\documentclass{aa}
\usepackage{graphicx}
\usepackage[varg]{txfonts}
\usepackage{natbib}
\usepackage{gensymb}
\usepackage{multirow}
\usepackage{color}
\usepackage[normalem]{ulem}

\bibpunct{(}{)}{;}{a}{}{,} 

\begin{document}

\title{Blanco DECam Bulge Survey (BDBS) III: A new view of the double red clump in the Milky Way bulge through luminosity and color distribution}


\author{Dongwook Lim\inst{1}
  \and Andreas J. Koch-Hansen\inst{1}
  \and Chul Chung\inst{2}
  \and Christian I. Johnson\inst{3}
  \and Andrea Kunder\inst{4}
  \and Iulia T. Simion\inst{5}
  \and R. Michael Rich\inst{6}
  \and William I. Clarkson\inst{7}
  \and Catherine A. Pilachowski\inst{8} 
  \and Scott Michael\inst{8}
  \and A. Katherina Vivas\inst{9}
  \and Michael D. Young\inst{10}
  }

\authorrunning{D. Lim et al.}
\titlerunning{Color distributions of the double red clump in the Milky Way bulge}


\institute{Zentrum f\"ur Astronomie der Universit\"at Heidelberg, Astronomisches Rechen-Institut, M\"onchhofstr. 12-14, 69120 Heidelberg, Germany, 
  \email{dongwook.lim@uni-heidelberg.de}
  \and Center for Galaxy Evolution Research \& Department of Astronomy, Yonsei University, Seoul 03722, Republic of Korea
  \and Space Telescope Science Institute, 3700 San Martin Drive, Baltimore, MD 21218, USA
  \and College of Arts \& Sciences, St Martin's University, Ernsdor Center 130 5000 Abbey Way SE Lacey, WA 98503, USA
  \and Key Laboratory for Research in Galaxies and Cosmology, Shanghai Astronomical Observatory, 80 Nandan Road, Shanghai 20030, China
  \and Department of Physics \& Astronomy, University of California Los Angeles, 430 Portola Plaza, Box 951547, Los Angeles, CA 90095-157, USA
  \and Department of Natural Sciences, University of Michigan-Dearborn, 4901 Evergreen Rd, Dearborn, MI 48128, USA
  \and Indiana University Department of Astronomy, SW319, 727 E 3rd Street, Bloomington, IN 47405 USA
  \and Cerro Tololo Inter-American Observatory, NSF's National Optical-Infrared Astronomy Research Laboratory, Casilla 603, La Serena, Chile
  \and Indiana University, University Information Technology Services, CIB 2709 E 10th Street, Bloomington, IN 47401 USA
  }

\date{Received November 21, 2020 / Accepted December 30, 2020}

\abstract {
Red clump (RC) stars are one of the best stellar tracers of the structure of the Milky Way (MW) bulge. Here we report a new view of the double RC through luminosity and color distributions of RC stars in nine bulge fields ($l$ = 0.0$\degree$, $\pm$4.5$\degree$; $b$ = -6.0$\degree$, -7.5$\degree$, -9.0$\degree$) from the Blanco DECam Bulge Survey (BDBS), which covers near-ultraviolet to near-infrared bandpasses. The bright and faint RCs show contrasting distributions in ($u-g$)$_{0}$ and ($u-i$)$_{0}$ colors but similar distributions in ($J-K_{s}$)$_{0}$ with a variation depending on the Galactic longitude, where the bright RC is typically redder than the faint RC. In particular, the RC stars are clearly divided into the bluer and redder populations when using the ($u-g$)$_{0}$ color (($u-g$)$_{0}$ $<$ 2.5 for the bluer RC; ($u-g$)$_{0}$ $\ge$ 2.5 for the redder RC). The bluer stars show a single clump on the faint RC regime, whereas the redder stars form double clumps on both the bright and faint RCs. The bright clump of the redder stars is dominant in the positive longitude fields, while the faint clump of those red stars is significant at negative longitudes. We also confirm that the bluer and redder stars have different peak metallicity through comparison with spectroscopy ($\Delta$[Fe/H] $\sim$ 0.45 dex). Therefore, our results support a scenario whereby the MW bulge is composed of a spheroid of metal-poor stars and a boxy/peanut shape (X-shape) predominantly made up of  metal-rich stars.
} 

\maketitle


\section{Introduction} \label{sec:intro}

The double red clump (RC) observed in the high-latitude fields of the Milky Way (MW) bulge is an essential feature for understanding the nature of the bulge. \citet{McWilliam2010} and \citet{Nataf2010} simultaneously reported that  the RC stars in the bulge can be divided into two groups, namely bright and faint RCs, from Two Micron All Sky Survey (2MASS) and Optical Gravitational Lensing Experiment (OGLE)-III surveys. The RC stars have long been used as a standard candle in order to examine the distance and structure of the Local Universe, and in particular the Galactic bulge \citep[e.g.,][]{Stanek1994, Rattenbury2007, Wegg2015}. Furthermore, the presence of the double RC can be taken as evidence for an X-shaped feature where the bright and faint RCs are placed on the near- and far-side arms of this structure \citep[see][]{Ness2012, Wegg2013}. We note that an X-shaped structure has been considered as a part of the boxy/peanut bulge and is indeed observed in several external galaxies \citep{Bureau2006, Buta2015, Gonzalez2016}. This X-shaped scenario for the origin of the double RC is based on the two main characteristics of the RC stars: firstly, stars in the bright and faint RCs show an almost  identical distribution in the ($J-K$) and ($V-I$) colors, which suggests a negligible difference in metallicity. Secondly, the peak magnitudes of the bright and faint RCs are almost constant regardless of Galactic longitude, while the population ratio between the two RCs changes significantly, which is inconsistent with the expected influence of the Galactic bar component \citep{McWilliam2010}. 

Multiple chemical populations may complicate the interpretation of the RC apparent magnitude distribution, as varying chemical compositions can change the RC absolute magnitude --this is generally observed in globular clusters (GCs). It is well established that GCs contain more than two stellar populations: the later-generation stars are more enhanced in He and certain light elements, such as N, Na, and Al, than the earlier-generation stars, and are depleted in others like C, O, and Mg \citep{Gratton2012, Bastian2018}. In the same vein, \citet{Lee2015} demonstrated  through population synthesis modeling that He-enhanced later-generation stars would be placed on the bright RC (bRC) regime, while the He-normal first-generation stars are mainly located on the faint RC (fRC) regime. Thus, the multiple population phenomenon may exist in the bulge, where the combined effects of different metallicity and He abundance can cause the RC stars to have different intrinsic luminosity \citep[see][]{Joo2017}. Recently, \citet{Lee2018} and \citet{Lim2021} reported  a difference in CN band strength and [Fe/H] between stars in the bright and faint RCs as supporting evidence of this scenario. These chemical properties of the double RC show some similarities with those observed among multiple stellar populations in the peculiar GC Terzan~5 \citep{Origlia2011}. Unlike the X-shaped scenario, the multiple population scenario does not require any distance difference between the bright and faint RCs. In particular, the multiple population scenario favors the ``classical bulge'' model, whereas the X-shaped scenario follows the ``pseudo bulge'' model. Therefore, investigating the properties of RC stars is crucial for understanding the structure of the MW bulge \citep[see also][]{Athanassoula2005, Nataf2017}. 

On the other hand, the MW bulge is known to have more than two stellar components with potentially different metallicity and kinematics, such as velocity dispersion and rotation curve \citep[e.g.,][]{Babusiaux2010, Ness2013, Zoccali2017,Clarkson2018}. The stars in the bulge can be mainly divided into the metal-poor ([Fe/H] $\sim$ $-$0.25 dex) and metal-rich ([Fe/H] $\sim$ +0.15 dex) components, although the metallicity criterion varies depending on the study. The majority of the metal-poor component shows a spheroid shape with roughly constant velocity dispersion with Galactic latitude (although we note that a barred distribution can also be seen in metal-poor stars; \citealt{Kunder2020}), whereas the metal-rich component forms a boxy/peanut shape with a steeper gradient of velocity dispersion with latitude. Thus, it appears that the MW bulge region contains both metal-poor stars characteristic of a classical bulge and metal-rich stars characteristic of a pseudo bulge, as well as the stellar populations of the inner halo and discs (\citealt{Rojas-Arriagada2014, Koch2016, Kunder2016, Savino2020}; see also \citealt{Kunder2020}).

Most studies of the MW bulge have been carried out using spectroscopic or photometric observations covering optical to near-infrared (NIR) bands. In this regard, the Blanco DECam Bulge Survey \citep[BDBS;][]{Rich2020} operating in the near-ultraviolet (NUV) to NIR has opened up new opportunities for studying the structure of the bulge. In particular, the NUV and optical colors of stars are more sensitive to stellar metallicity, age, and chemical composition of light elements. The BDBS is the Rubin Observatory Legacy Survey of Space and Time (LSST) precursor program covering $\sim$200 square degrees of the Galactic bulge at $-$11$\degree$ $<$ $l$ $<$ +11$\degree$ and $-$13$\degree$ $<$ $b$ $<$ $-$2$\degree$ and was performed from 2013 to 2014 using the Dark Energy Camera (DECam) at the CTIO-4m telescope with $ugrizY$ filters. The main goal of the BDBS is to produce an optical multi-band map of the southern Galactic bulge. A more detailed description of the BDBS is presented in \citet{Rich2020} and \citet{Johnson2020}. 

Here, we investigate the luminosity and color distributions of RC stars in various bulge fields taking advantage of metallicity-sensitive BDBS photometry with NIR photometry of the Vista Variables in the Via Lactea \citep[VVV;][]{Minniti2010}, complemented by parallax information from the second {\em Gaia} data release (DR2, \citealt{GaiaCollaboration2018}). The current paper is organized as follows. In Section~\ref{sec:data}, we describe the data-selection process. The luminosity and color distributions of RC stars are presented in Section~\ref{sec:color}, while our results are compared with spectroscopy in Section~\ref{sec:spec}. Finally, our conclusion and discussion for the MW bulge drawn from the RC stars are given in Section~\ref{sec:discussion}. 


\section{Data selection} \label{sec:data}

In order to investigate the color and luminosity distributions of RC stars, we obtained $ugrizY$ magnitudes from the BDBS for stars within circles of  1$\degree$ diameter around nine different fields of the bulge at $l$ = 0.0$\degree$, $\pm$4.5$\degree$ and $b$ = -6.0$\degree$, -7.5$\degree$, and -9.0$\degree$. We first focus on the central field at ($l$, $b$) = (0.0$\degree$, -7.5$\degree$) where the double RC is distinctly observed, and then select the near-fields toward increasing or decreasing longitude and latitude to examine the trends of the RC depending on Galactic position. Here, we note that the double RC feature is most prominently observed in the high-latitude fields (|$b$| $\ge$  6$\degree$; see \citealt{McWilliam2010, Nataf2010}). There is only one GC in the selected areas and contamination is minimized by excluding stars within one half-light radius of that one, namely NGC~6558. We then selected the best sample of stars from the BDBS data by applying the criteria for measurement error, observing count and quality flags for each band: error1 $\le$ error2, count $\ge$ 2, error{\_}flag $\le$ 1, sky{\_}flag $\le$ 2, shape{\_}flag $\le$ 1. Here, error1 is calculated from the weighted flux, and error2 is the magnitude error added in quadrature for each exposure. Three quality flags indicate the number of standard deviations away from the mean of error, sky, and chi values, respectively. A detailed description of the data-reduction process of the BDBS and its quality flags can be found in \citet{Johnson2020}. However, in the case of ($l$, $b$) = (+4.5$\degree$, -9.0$\degree$) field,  the bulk of bright stars ($K_{s_{0}}$ $<$ 12.5) is excluded with these criteria because of the low number of observations at the edge of the survey \citep[see Figure~1 of][]{Rich2020}. Therefore, we applied more relaxed criteria for this particular field, namely error1 $\le$ error2, error{\_}flag $\le$ 2, and sky{\_}flag $\le$ 2.
 
\begin{table}
\tiny
\caption{Number of stars in each field}
\label{tab:number} 
\centering                                    
\begin{tabular}{c c c c}          
\hline\hline                        
N$_{total}$                     & \multirow{2}{*}{$l$ = +4.5$\degree$} & \multirow{2}{*}{$l$ = 0.0$\degree$} & \multirow{2}{*}{$l$ = -4.5$\degree$} \\
(N$_{RGB}$; N$_{RC})$   & & & \\
\hline                                   
\multirow{2}{*}{$b$ = -6.0$\degree$}    & 177,214               & 93,887                         & 90,713                        \\
                                                        & (29,648; 18,925)         & (29,373; 19,817)      & (27,383; 18,428)      \\
\hline
\multirow{2}{*}{$b$ = -7.5$\degree$}    & 154,298               & 72,895                         & 125,829                       \\
                                                        & (16,352; 9,962)         & (8,290; 4,908)        & (11,500; 6,687)       \\
\hline
\multirow{2}{*}{$b$ = -9.0$\degree$}    & 82,176                        & 87,258                  & 114,217                       \\
                                                        & (4,496; 2313)         & (8,176; 4,846)  & (5,297; 2,588)        \\
\hline                                             
\end{tabular}
\end{table}

\begin{figure*}
\centering
   \includegraphics[width=0.96\textwidth]{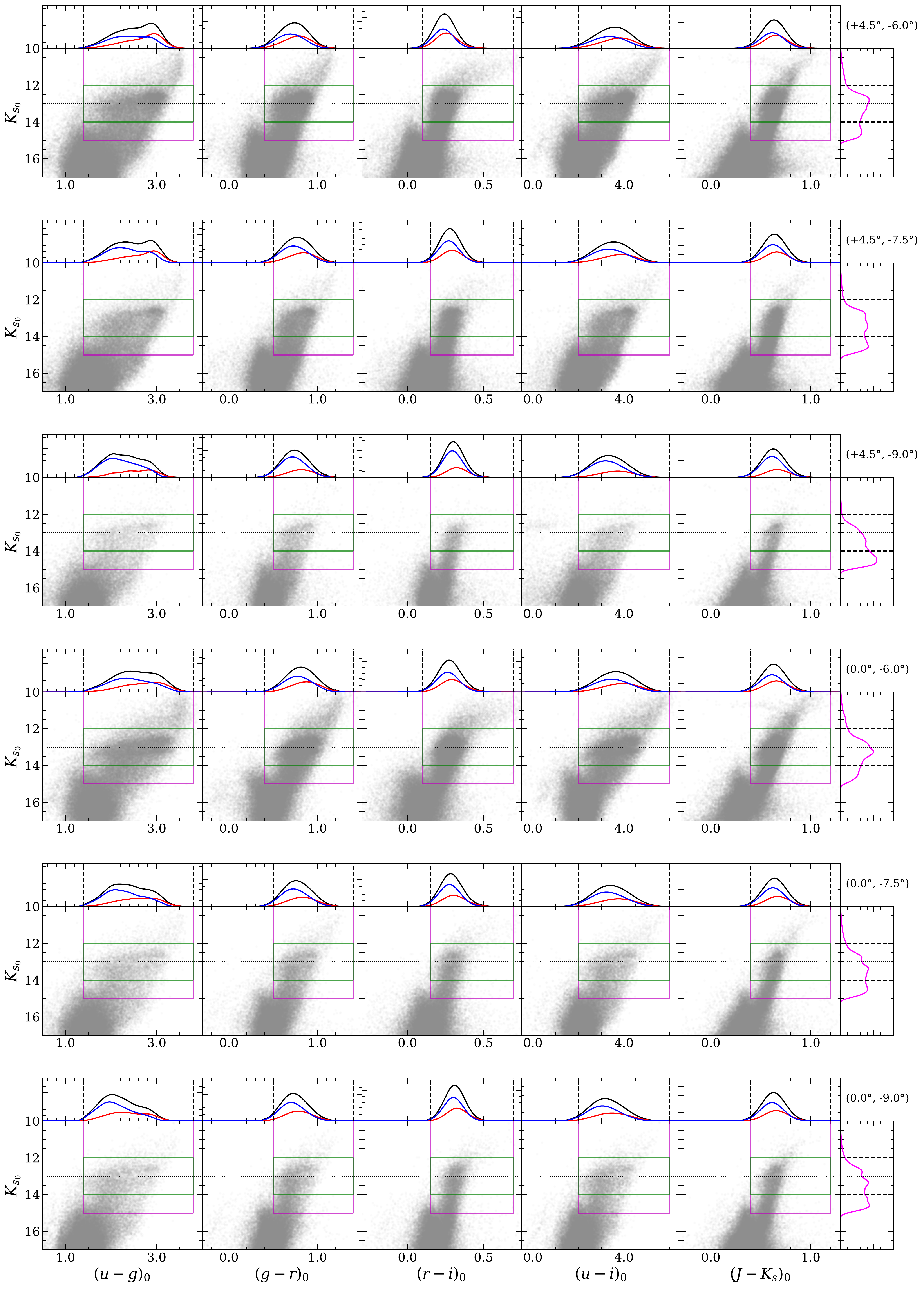}
\end{figure*}

\begin{figure*}
\centering
   \includegraphics[width=0.96\textwidth]{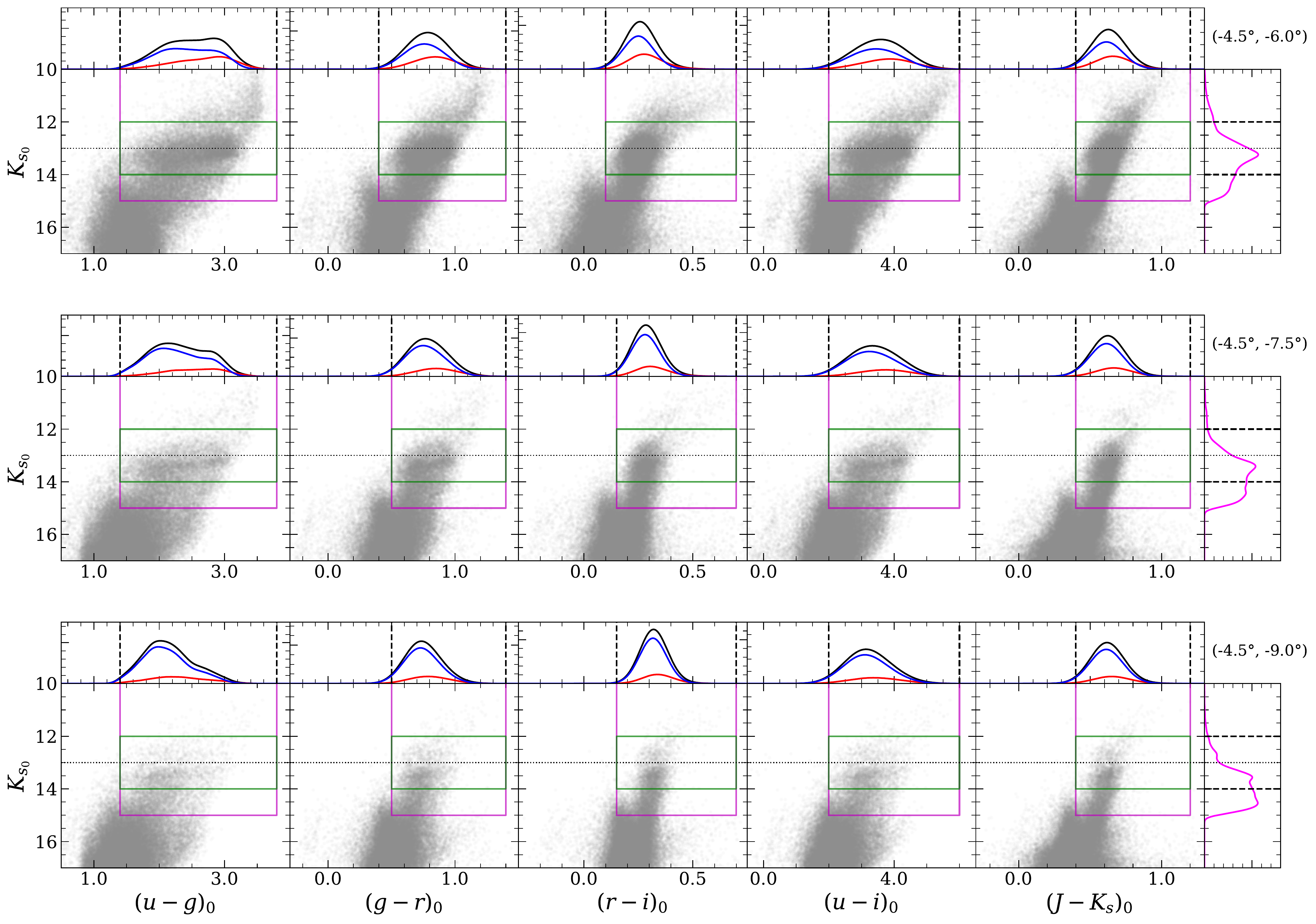}
     \caption{Dereddened CMDs for stars within a circle of 1$\degree$ diameter around nine different fields of the bulge in the ($K_{s}$, $u-g$)$_{0}$, ($K_{s}$, $g-r$)$_{0}$, ($K_{s}$, $r-i$)$_{0}$, ($K_{s}$, $u-i$)$_{0}$, and ($K_{s}$, $J-K_{s}$)$_{0}$ planes using the BDBS and the VVV data. 
     The Galactic position ($l$, $b$) of each field is listed in the upper right corner. 
     The green and purple boxes indicate the selection criteria for the RC and RGB stars, respectively, and the horizontal dotted line in each CMD divides the bright and faint RCs at 13.0 $K_{s_{0}}$~mag. 
     We also plot the luminosity function of the RGB stars for each field (rightmost panels in each field), and color distribution for the stars in bRC (red), fRC (blue), and RGB (black) regimes, respectively, on the top of each CMD.
     The presence of double RCs is particularly prominent in the fields of $l$ = 0.0$\degree$, while the bRC (fRC) is dominant at the positive (negative) longitude fields ($l$ = $\pm$4.5$\degree$). The stars in the bright and faint RC show a different distribution in ($u-g$)$_{0}$ and ($u-i$)$_{0}$ colors, but similar distribution in ($r-i$)$_{0}$ and ($J-K_{s}$)$_{0}$ colors.}
     \label{fig:CMD}
\end{figure*}

In addition, we cross-matched the BDBS data with the Gaia DR2 and the VVV DR2 NIR photometry using the CDS X-Match service from {\em TOPCAT} \citep{Taylor2005}. We note again that the two major obstacles in the study of the bulge are the contamination of stars from other stellar components towards the Galactic plane, such as disk and inner halo, and the high interstellar reddening toward the bulge. Although the accuracy of the Gaia parallax ($\varpi$) measurements is insufficient to fully disentangle the bulge stars from the Galactic disk, we only used the stars within the range of -0.2 $<$ $\varpi$ $<$ 0.4 and -2.0 $<$ relative parallax error ($\varpi${/}$\sigma_{\varpi}$) $<$ 4.0 for this work at least to exclude nearby stars. In each field, about 45\% of stars are excluded by this selection criterion and the majority of them seem to be main sequence stars belonging to the thin and thick disks \citep[see Figure~7 of][]{Rich2020}. We note that this selection procedure by parallax does not significantly affect our results, because, in general, about 80\% of RGB and 85\% of RC stars still remain in the samples. The NIR photometry obtained from the VVV is used for identifying RC stars because these bands are less sensitive to the reddening effect. The double RC feature is indeed most prominently observed in the NIR bands \citep[see][]{McWilliam2010, Wegg2013}.  

Following \citet{Johnson2020}, we derived the reddening corrected magnitudes using the extinction map from \citet{Simion2017} and the extinction coefficients of \citet{Schlafly2011} for the $u$-band, \citet{Green2018} for the $grizY$-bands, \citet{Alonso-Garcia2017} for the $JHK$-bands, and \citet{Casagrande2018} for the Gaia photometry. Figure~\ref{fig:CMD} shows color--magnitude diagrams (CMDs) for our sample stars from the BDBS within our parallax range in the nine Galactic bulge fields, showcasing the ($u-g$)$_{0}$, ($g-r$)$_{0}$, ($r-i$)$_{0}$, ($u-i$)$_{0}$, and ($J-K_{s}$)$_{0}$ colors. The number of stars in each field is listed in Table~\ref{tab:number}. 


\section{Color distribution of RC stars} \label{sec:color}

First of all, we identify the red giant branch (RGB) and RC stars from the CMDs in Figure~\ref{fig:CMD}, which simultaneously fall on the specific regions on each CMD  (purple boxes for RGB stars; green boxes for RC stars). We note that we slightly adapted the color criteria with Galactic latitude, while the magnitude ranges were kept identical for all fields as 10.0 $<$ $K_{s_{0}}$~mag $<$ 15.0 for RGB stars and 12.0 $<$ $K_{s_{0}}$~mag $<$ 14.0 for RC stars. All stars in the RC regime are also included in the RGB group because the RC and RGB stars could not be distinguished on the CMD. Table~\ref{tab:number} shows the number of stars in the RGB and RC regimes for each field. 

The rightmost  panels of Figure~\ref{fig:CMD} present the luminosity function of RGB stars. These show a bimodal distribution at the magnitude range of the RC, which can be divided into the bright and faint RCs at $K_{s_{0}}$~mag = 13.0 (horizontal dotted line). This double RC feature is particularly apparent in the fields of $l$ = 0.0$\degree$ and becomes significant with increasing latitude. 
\footnote{It is important to note that, besides RC stars, stars in the evolutionary stage of the red giant branch bump (RGBB) are also embedded in the luminosity function of RGB stars. In particular, the RGBB stars of the bRC, corresponding to $\sim$25\% of the number counts of the bRC stars, might be placed within the similar magnitude range of the fRC \citep[see][]{Nataf2014}. However, in  this study  the contamination by RGBB stars was not taken into account because it is similar in every bulge field.}
In addition, the bRC is more dominant than the fRC in the positive longitude fields ($l$ = +4.5$\degree$), whereas the fRC stars are more abundant at negative longitudes ($l$ = -4.5$\degree$). These trends of the double RC depending on the Galactic position are identical to the previous reports by \citet{McWilliam2010} and \citet{Nataf2015}. 

In order to examine the color distribution of the double RCs, we divide RC stars into bright and faint RCs (12.0 $<$ $K_{s_{0}}$ $\le$ 13.0 for bRC; 13.0 $<$ $K_{s_{0}}$ $\le$ 14.0 for fRC) and then draw the histograms of each color in the top panels of the CMD in Figure~\ref{fig:CMD} (blue for fRC; red for bRC; black for RGB). It is expected that the bRC is typically redder than the fRC in every histogram because the RGB stars become redder with increasing luminosity. Nevertheless, the differences in ($u-g$)$_{0}$ and ($u-i$)$_{0}$ between the two RCs are more obvious than those in other colors. In particular, in the fields of ($l$, $b$) = (0.0$\degree$, -7.5$\degree$), it appears that both the bRC and fRC show the bimodal distribution in ($u-g$)$_{0}$ with a stronger redder peak in the bRC and a stronger bluer peak in the fRC. Thus, these CMDs and histograms of the BDBS data suggest a possible difference in color distribution between the bright and faint RCs. 

\begin{figure}
\centering
   \includegraphics[width=0.47\textwidth]{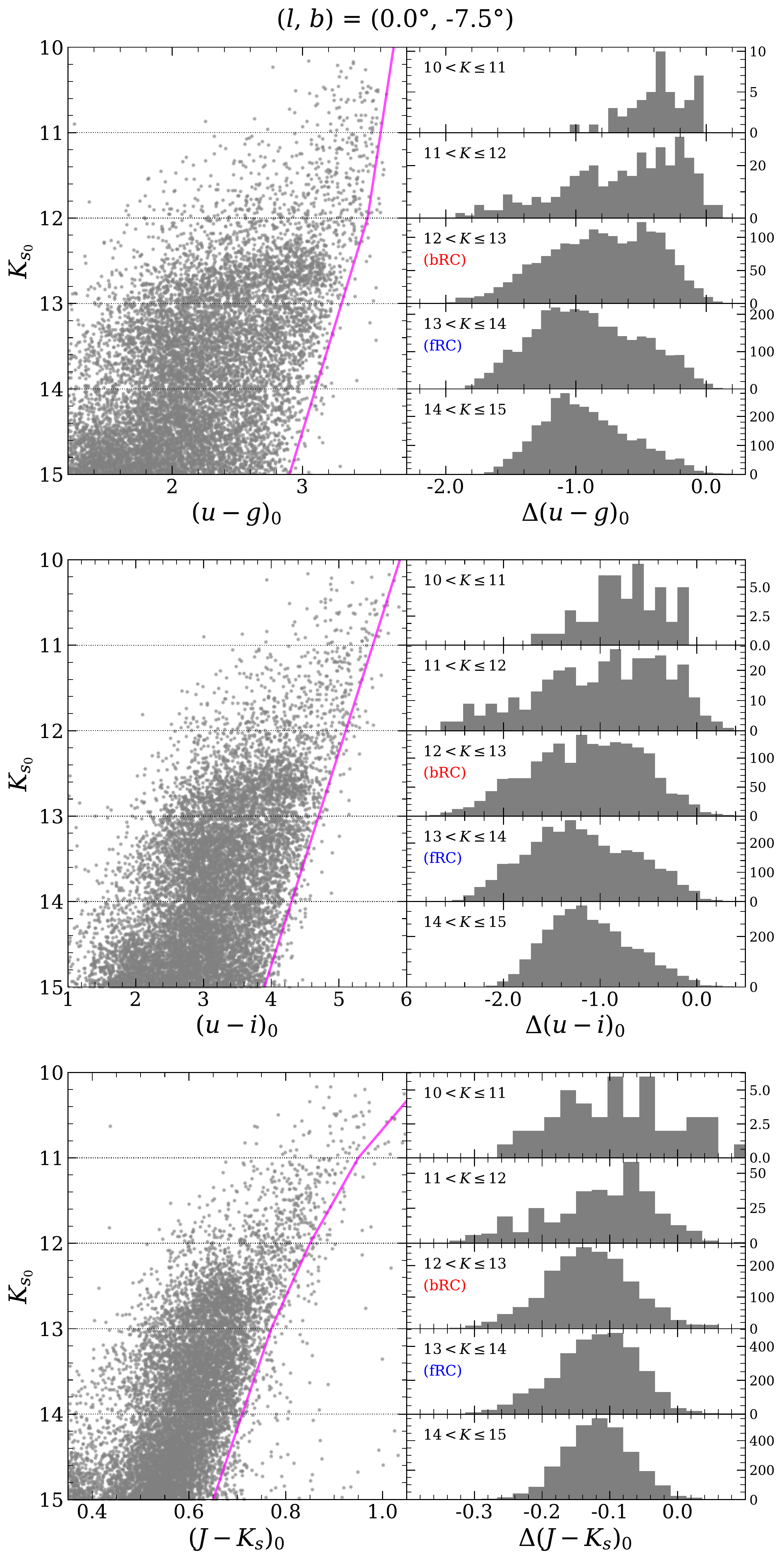}
     \caption{Color--magnitude diagrams and histograms of $\Delta$($u-g$)$_{0}$, $\Delta$($u-i$)$_{0}$, and $\Delta$($J-K_{s}$)$_{0}$ for stars in the field of ($l$, $b$) = (0.0$\degree$, -7.5$\degree$). 
     The $\Delta$-colors are derived as the difference between the original color and the fiducial line (purple lines in the left panels), and the histogram is respectively drawn for stars within the 1.0 mag range from 10.0 to 15.0 mag in $K_{s_{0}}$-band. 
     Both the bRC and fRC stars show bimodal distributions in $\Delta$($u-g$)$_{0}$ and $\Delta$($u-i$)$_{0}$. 
     However, the majority of the bRC is in the redder side, whereas that of the fRC is in the bluer side.
     In contrast, the bRC and fRC show similar distributions in the $\Delta$($J-K_{s}$)$_{0}$.
     }
     \label{fig:hist}
\end{figure}

However, it is necessary to confirm that the difference in color distribution between the bright and faint RCs is not due to the general trend on the RGB. Therefore, we determine the ``delta colors'' for stars in the field of ($l$, $b$) = (0.0$\degree$, -7.5$\degree$) as the horizontal distance from the fiducial line (purple lines in Figure~\ref{fig:hist}), which is visually defined as the right edge of the RGB similar to \citet{Lee2013}. Figure~\ref{fig:hist} shows CMDs in the ($K_{s}$, $u-g$)$_{0}$, ($K_{s}$, $u-i$)$_{0}$, and ($K_{s}$, $J-K_{s}$)$_{0}$ planes, together with histograms of the respective delta colors for stars in each bin of 1.0 mag from 10.0 to 15.0 in $K_{s_{0}}$-band. The histograms indicate the different patterns in the $\Delta$($u-g$)$_{0}$ and $\Delta$($u-i$)$_{0}$ between the bRC and fRC regimes. Although both the bRC and fRC have two peaks at the bluer and redder colors, the majority of the bRC is in the redder peak while that of the fRC is in the bluer peak. Thus, the bRC stars are generally redder than the fRC stars in the ($u-g$)$_{0}$ and ($u-i$)$_{0}$ colors regardless of the trend of the RGB. We note that the bimodal distribution of the bRC and fRC in ($u-i$)$_{0}$ color has already been reported in the field of ($l$, $b$) = (+1$\degree$, -8$\degree$) from the previous BDBS study (see Figure~17 of \citealt{Johnson2020}). In the case of $\Delta$($J-K_{s}$)$_{0}$, the two RCs show a similar distribution, which is consistent with the earlier finding by \citet{McWilliam2010}. The $\Delta$($J-K_{s}$)$_{0}$ color of the bRC is even somewhat bluer than that of the fRC in contrast to the cases of $\Delta$($u-g$)$_{0}$ and $\Delta$($u-i$)$_{0}$. This discrepancy implies that the NUV and optical photometry of the BDBS is highly powerful and provides a new view of the double RC in the bulge. 

\begin{figure*}
\centering
   \includegraphics[width=1.0\textwidth]{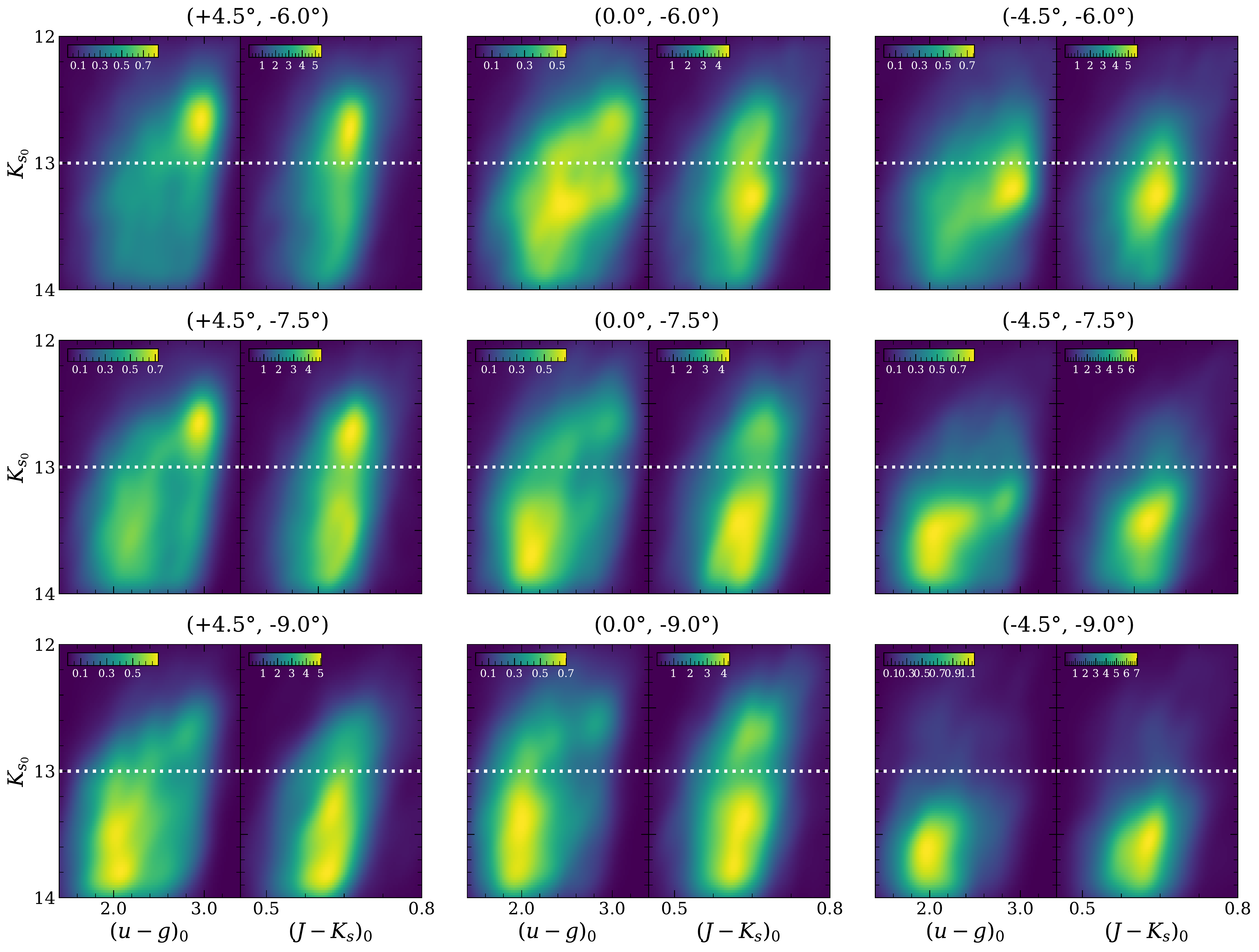}
     \caption{Density maps of stars in the RC regime for nine Galactic fields in the ($K_{s}$, $u-g$)$_{0}$ and ($K_{s}$, $J-K_{s}$)$_{0}$ planes. 
     The horizontal dotted line indicates 13.0 mag in the $K_{s_{0}}$-band, which divides the bright and faint RCs. 
     While the two RCs have similar ($J-K_{s}$)$_{0}$ color, they show contrasting distribution in the ($K_{s}$, $u-g$)$_{0}$ plane with large variations depending on the Galactic position.
     In general, the bright stars form a redder clump than the faint stars in the ($u-g$)$_{0}$ color (see text for details). 
     }
     \label{fig:density}
\end{figure*}

A more detailed comparison of the color and luminosity distributions between the bright and faint RCs, as well as the longitude and latitude dependence of the double RC, is presented in Figure~\ref{fig:density}, which shows density maps of the stars on the RC in the ($K_{s}$, $u-g$)$_{0}$ and ($K_{s}$, $J-K_{s}$)$_{0}$ planes for all fields used in this study. As has been shown before, the double RC is prominently observed in the ($K_{s}$, $J-K_{s}$)$_{0}$ CMD, particularly at the fields of $l$ = 0.0$\degree$. The bright and faint RCs are clearly separated in $K_{s_{0}}$-magnitude with a similar color of ($J-K_{s}$)$_{0}$. The change in the fraction of the bright and faint RC stars depending on the Galactic longitude is also well demonstrated in the ($K_{s}$, $J-K_{s}$)$_{0}$ plane of Figure~\ref{fig:density}. For instance, the bRC is more significant than the fRC in the positive longitude fields ($l$ = +4.5$\degree$), but this trend is reversed at the negative longitude fields ($l$ = -4.5$\degree$). However, in the field of ($l$, $b$) = (+4.5$\degree$, -9.0$\degree$),  the dominance of the bRC is less clear compared to other fields of ($l$, $b$) = (+4.5$\degree$, -6.0$\degree$) and (+4.5$\degree$, -7.5$\degree$). It is probably due to the lack of bright stars in this field during the sample-selection procedure, although we applied the relaxed criteria for this field (see Section~\ref{sec:data} and Figure~\ref{fig:CMD}).

Interestingly, the bright and faint RCs show contrasting distributions in ($u-g$)$_{0}$ color. The fRC stars are concentrated in the bluer regime at around ($u-g$)$_{0}$ $\sim$ 2.0 for all fields. Although the bRC stars are mainly placed in the redder regime at ($u-g$)$_{0}$ $\sim$ 3.0, their distribution patterns are varied with Galactic position. In particular, while the bRC stars show a redder clump with a tail toward  bluer colors in the fields at $l$ = 0.0$\degree$ and +4.5$\degree$, the fainter and redder clump is shown in the fields of $l$ = -4.5$\degree$ at around a $K_{s_{0}}$~mag of 13.3  and 2.9 in ($u-g$)$_{0}$, instead of a distinct bRC. In addition, closer inspection of ($l$, $b$) = (+4.5$\degree$, -7.5$\degree$) and (0.0$\degree$, -7.5$\degree$) fields reveals that the RGB stars can also be divided into the bluer and redder branches in ($u-g$)$_{0}$ color, in addition to the RCs (see also Figure~\ref{app:fig:rgb}). Additional density maps using other color combinations are shown in Appendix~\ref{sec:appendix}.

\begin{figure*}
\centering
   \includegraphics[width=0.6\textwidth]{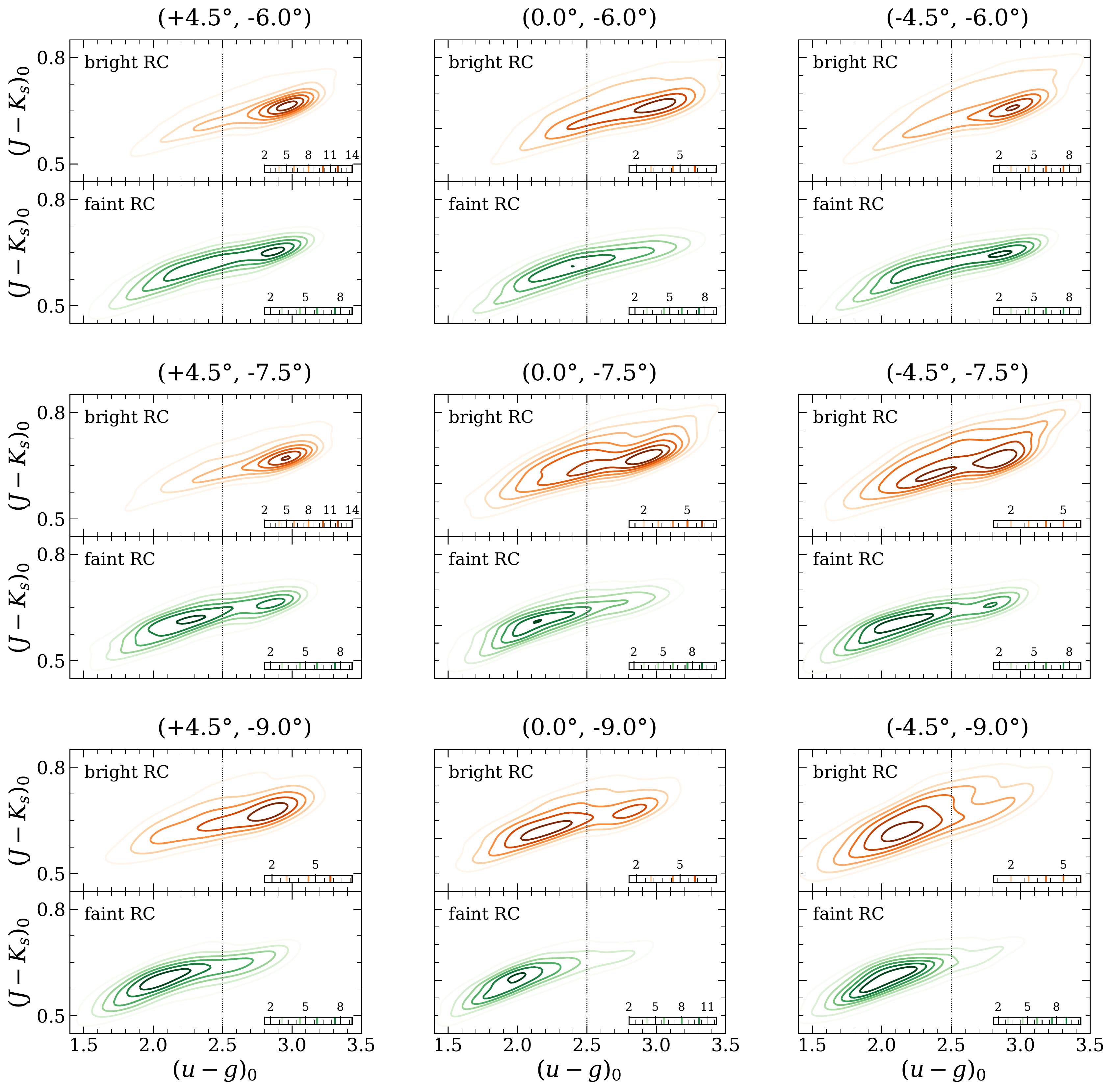}
     \caption{Density contours of stars in the bright and faint RC regimes, respectively, in the ($J-K_{s}$, $u-g$)$_{0}$ plane.
     Stars on  both RCs can be divided into two subgroups with different ($u-g$)$_{0}$ and similar ($J-K_{s}$)$_{0}$ in most fields. 
     The vertical dotted line indicates ($u-g$)$_{0}$ = 2.5, where we divide the sample into  bluer and redder RC stars.
     }
     \label{fig:color_color}
\end{figure*}

In order to further examine the split of stars by color, we plot the color--color contours using ($u-g$)$_{0}$ and ($J-K_{s}$)$_{0}$ for stars on the bright and faint RCs, respectively, in Figure~\ref{fig:color_color}. As is clearly shown in the fields at $b$ = -7.5$\degree$, the stars in both the bRC and fRC are separated into two subgroups with different ($u-g$)$_{0}$, but similar ($J-K_{s}$)$_{0}$. Furthermore, a similar pattern is commonly observed in all fields for the bRC and fRC. We note that displaying the contours in two-color diagrams is more illustrative of the color distribution than the density map drawn from all stars in the RC regime of Figure~\ref{fig:density}. For instance, the presence of the redder clump in the bRC regime at ($l$, $b$) = (+4.5\degree$, -9.0\degree$), which is not evident in Figure~\ref{fig:density}, is distinct in this contour. Nevertheless, the separation by ($u-g$)$_{0}$ color is less clear in some fields, such as the bRC in the field at ($l$, $b$) = (+4.5\degree$, -6.0\degree$) and the fRC in ($l$, $b$) = (-4.5\degree$, -9.0\degree$), which is probably due to the relatively large difference in the number ratio between the bluer and redder subgroups. As the color of the RGB and RC stars is generally related to their metallicity, this subgrouping would imply the presence of two stellar populations with different metallicities in the outer MW bulge. For a detailed investigation of the color and magnitude distribution for these subgroups, we divide the stars in the RC regime into the bluer and redder RCs regardless of magnitude (($u-g$)$_{0}$ $<$ 2.5 for bluer RC stars; ($u-g$)$_{0}$ $\ge$ 2.5 for redder RC stars; vertical dotted line in Figure~\ref{fig:color_color}). 

\begin{figure*}
\centering
   \includegraphics[width=0.8\textwidth]{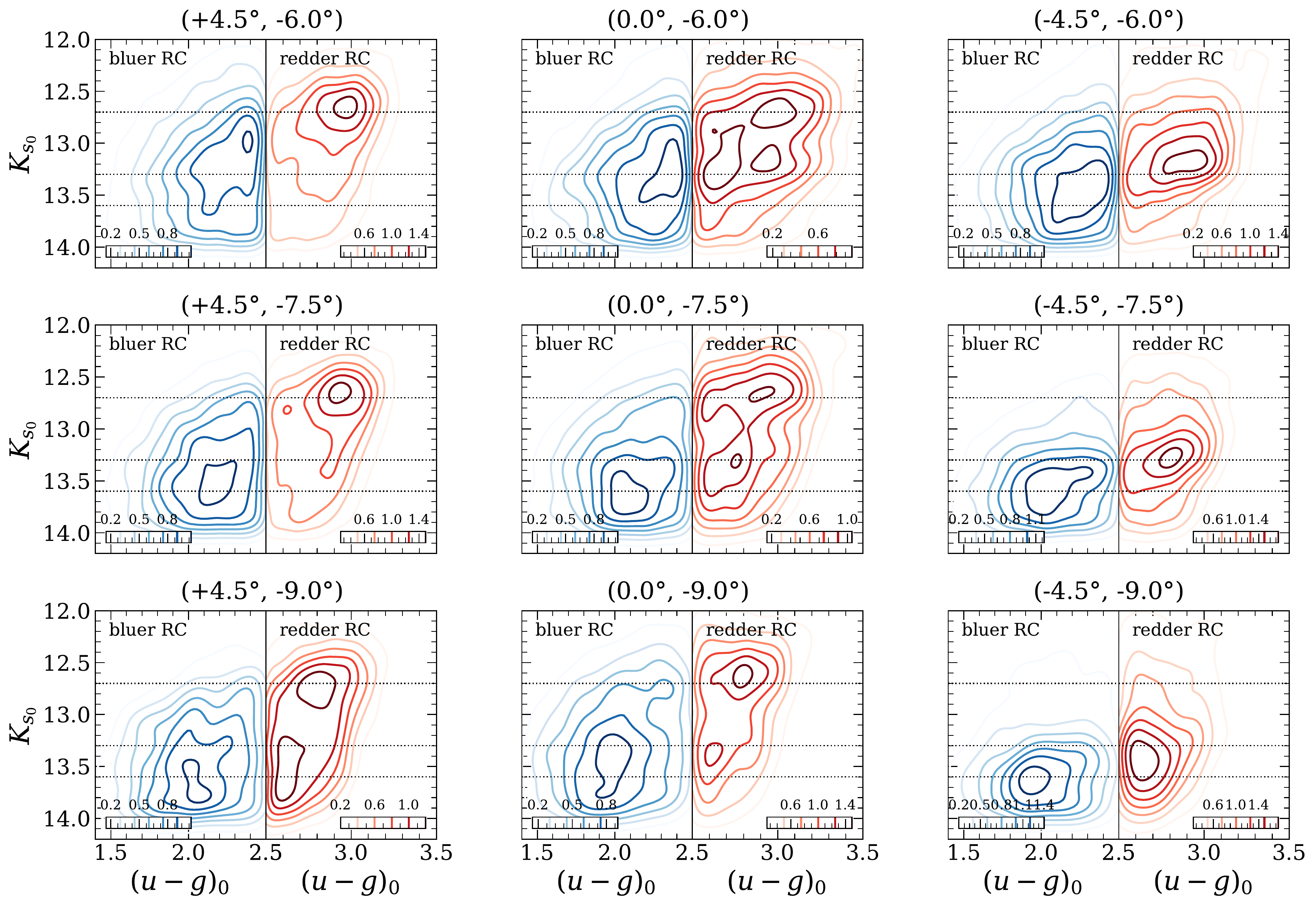}
     \caption{Density contours of stars in the RC regime with subgrouping by ($u-g$)$_{0}$ color, in the ($K_{s}$, $u-g$)$_{0}$ plane.  
     While the bluer RC stars show a single faint clump in all fields, the pattern of the redder RC stars varies with longitude and latitude. 
     In the redder RC regime, the bright (faint) clump is prominently observed in the fields of positive (negative) longitude, and both clumps are shown in the fields of $l$ = 0.0$\degree$.
     The peak magnitudes of these bright and faint clumps of redder stars are almost the same regardless of Galactic position.
     We note that the single clump of the bluer stars is fainter than the faint clump of the redder stars. 
     The horizontal dotted lines represent $K_{s_{0}}$~mag = 12.7, 13.3, and 13.6, respectively. 
     }
     \label{fig:contour_ug}
\end{figure*}

\begin{figure*}
\centering
   \includegraphics[width=0.55\textwidth]{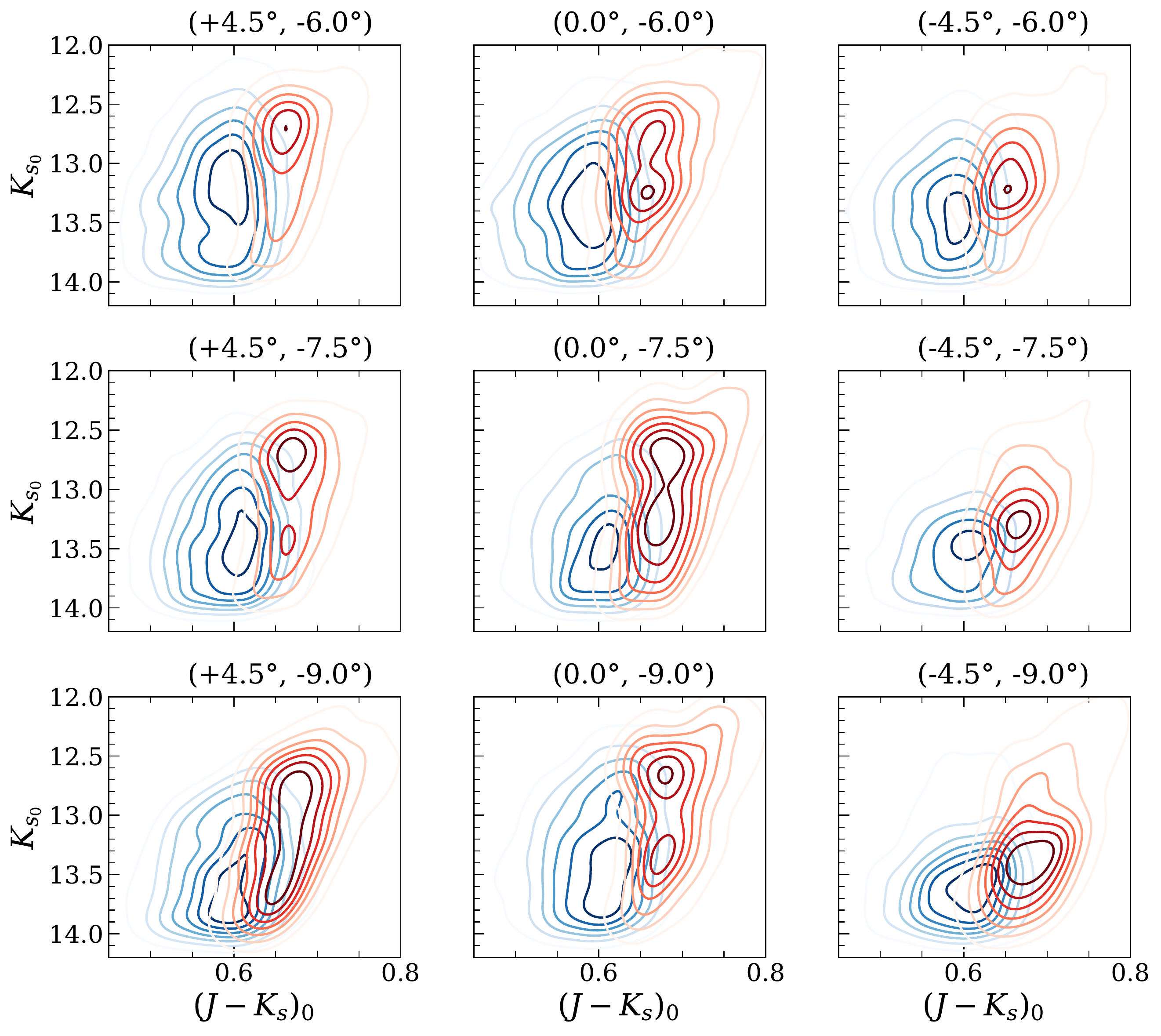}
     \caption{Same as Figure~\ref{fig:contour_ug} but for ($J-K_{s}$)$_{0}$ color. 
     The bluer and redder stars, divided by ($u-g$)$_{0}$, overlap in ($J-K_{s}$)$_{0}$ color.
     Thus, the distinct double RC observed in the ($K_{s}$, $J-K_{s}$)$_{0}$ CMD may be due to the the synergy of the faint clump of the bluer stars and the bright and faint clumps of the redder stars.
     }
     \label{fig:contour_jk}
\end{figure*}

\begin{figure*}
\centering
   \includegraphics[width=0.55\textwidth]{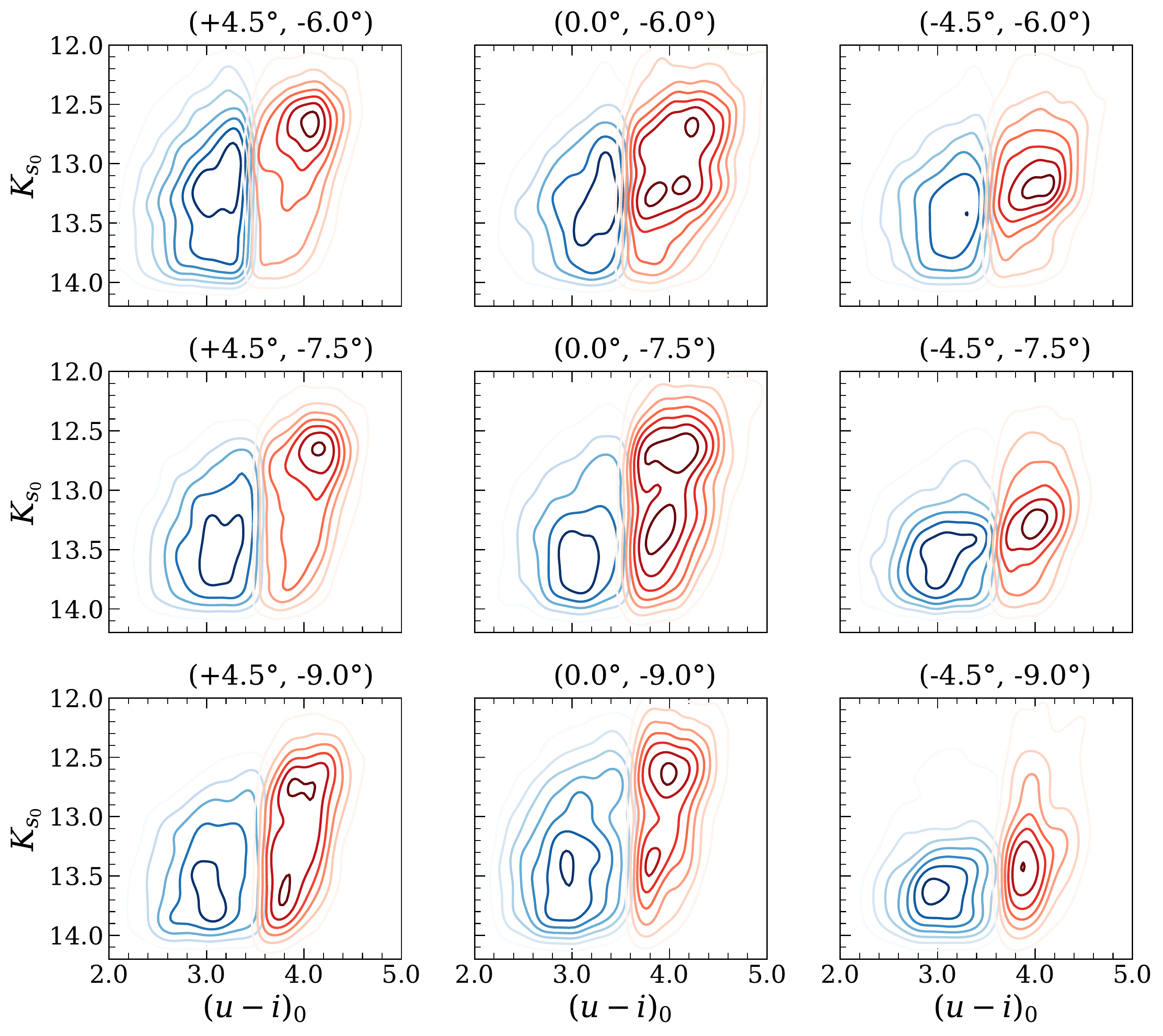}
     \caption{Same as Figure~\ref{fig:contour_ug} but for ($u-i$)$_{0}$ color.
     The blue and redder stars show almost the same pattern within the ($u-g$)$_{0}$ and ($u-i$)$_{0}$ colors.
     As the ($u-i$)$_{0}$ color correlates with metallicity \citep{Johnson2020}, the bluer and redder stars would have a significant difference in metallicity.  
     } 
     \label{fig:contour_ui}
\end{figure*}

\begin{figure*}
\centering
   \includegraphics[width=0.98\textwidth]{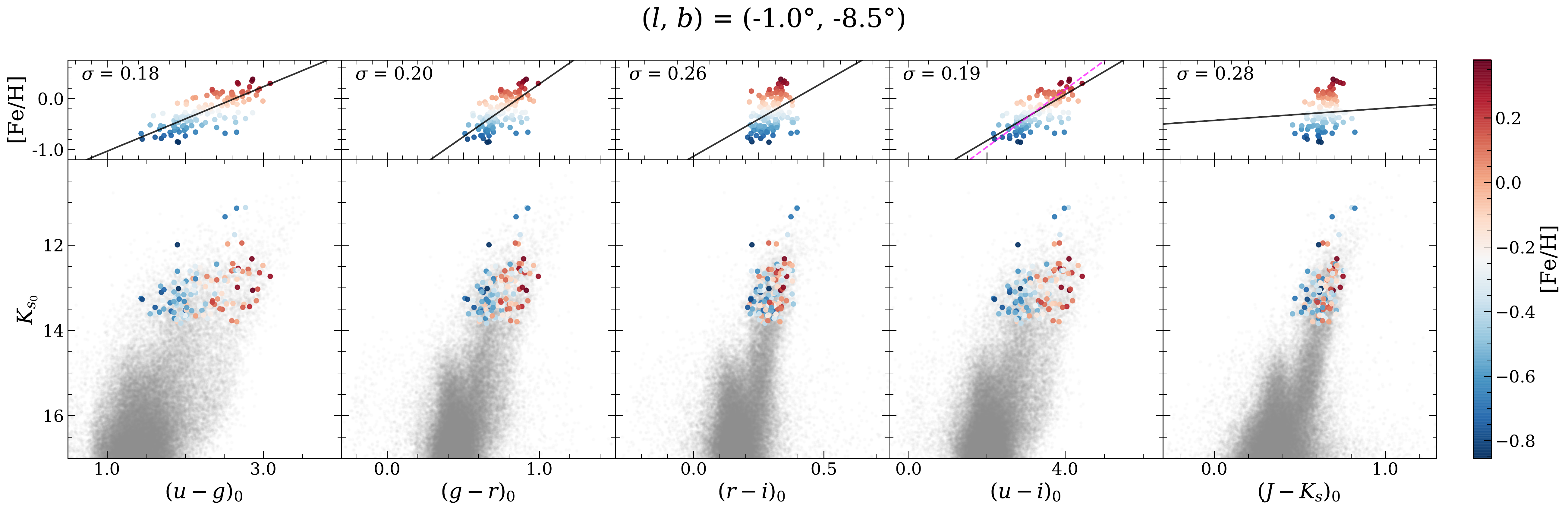}
     \caption{Color--magnitude diagrams for stars in the field at ($l$, $b$) = (-1.0$\degree$, -8.5$\degree$) together with  [Fe/H] abundances of stars obtained from spectroscopy \citep[data from][]{Lim2021}. 
     The colored circles indicate the metallicity of stars from metal-poor (blue) to metal-rich (red).
     The [Fe/H] abundances of stars are clearly enhanced with increasing color.
     In particular, the metallicity gradient is apparent in the ($u-g$)$_{0}$, ($g-r$)$_{0}$, and ($u-i$)$_{0}$ colors, but is indistinct in  ($r-i$)$_{0}$ and ($J-K_{s}$)$_{0}$. 
     The upper panels show color--metallicity relations (solid black lines) obtained from stars in 12.0 $\le$ $K_{s_{0}}$ $\le$ 14.0. 
     The standard deviation ($\sigma$) of the offset between the observed and fitted [Fe/H] values is indicated in the upper left corner.
     The [Fe/H] of stars is tightly correlated with ($u-g$)$_{0}$, ($g-r$)$_{0}$, and ($u-i$)$_{0}$ colors with a small standard deviation.
     We note that the dashed magenta line in the ($u-i$)$_{0}$ color represents the color--metallicity relation calculated in \citet{Johnson2020}. 
     }
     \label{fig:spec}
\end{figure*}

Figure~\ref{fig:contour_ug} shows the density contours in the ($K_{s}$, $u-g$)$_{0}$ plane with subgrouping of the bluer and redder stars in the RC regime.  First, when we examine the redder RC stars, the presence of the bright and faint clumps are clearly shown in the fields of $l$ = 0.0$\degree$, while the bright clump is more significant than the faint clump. The ratio of the faint clump to the bright clump becomes smaller at the higher latitude field, but the average magnitudes of these two clumps are almost constant at the $K_{s_{0}}$~mag of 12.7 for the bright clump and 13.3 for the faint clump regardless of latitude. In addition, the bright and faint clumps of the redder stars show a small difference in ($u-g$)$_{0}$ color ($\sim$0.2 dex) compared to that derived from all samples in the bRC and fRC ($\sim$ 1.0 dex; see Figure~\ref{fig:density}). The bright clump becomes predominant in the fields of positive longitude ($l$ = +4.5$\degree$), whereas the faint clump is significant in the fields of negative longitude ($l$ = -4.5$\degree$). It is important to note that the peak magnitudes of the bright and faint clumps are not changed with Galactic position. Thus, the two clumps of the redder RC stars show a negligible difference in color and a variation of ratio with Galactic longitude and latitude with constant magnitudes, which is identical to the properties of the double RCs expected from the X-shaped model \citep[e.g.,][see also Section~\ref{sec:intro}]{McWilliam2010}. In addition, amongst the fields of $l$ = 0.0$\degree$, the bright and faint clumps of redder RC stars are further apart in the higher latitude field. This also supports the X-shaped scenario for the double clumps in the redder RC regime. In contrast, the bluer RC stars only form a single extended clump on the fainter region at 13.3 $\sim$ 13.8 $K_{s_{0}}$~mag in all fields. In particular, the peak magnitude of this clump is even fainter than the faint clump of the redder RC stars. 

We also plot the density contours of the bluer and redder RC stars, divided by ($u-g$)$_{0}$ color, in the ($K_{s}$, $J-K_{s}$)$_{0}$ and ($K_{s}$, $u-i$)$_{0}$ planes in Figures~\ref{fig:contour_jk} and \ref{fig:contour_ui}. Although the blue and redder stars are similarly separated in the ($J-K_{s}$)$_{0}$ and ($u-i$)$_{0}$ colors, they mildly overlap in the ($K_{s}$, $J-K_{s}$)$_{0}$ plane. It therefore appears that the distinct double RC observed in the NIR photometry is composed of the bright clump of the redder stars and the faint clumps of both the redder and bluer stars. We note that similar distributions of bluer and redder stars are also observed in other color combinations (see Figures~\ref{app:fig:gi} and \ref{app:fig:ri}). In addition, as the ($u-i$)$_{0}$ color is tightly correlated with metallicity (see Figure~18 of \citealt{Johnson2020}), the difference in ($u-i$)$_{0}$ between the bluer and redder stars suggests a difference in metallicity, where the redder stars are more metal-rich. The fact that the double RC appears only amongst the redder stars while the bluer stars comprise a single clump corresponds to the current understanding of the MW bulge that the double RC feature is prominent among the metal-rich stars \citep{Ness2012, Rojas-Arriagada2014}. In the same vein, the difference in metallicity between stars in the bright and faint RCs reported by spectroscopic studies \citep[e.g.,][]{Uttenthaler2012} is also reasonable because the bluer (metal-poor) stars only form a single clump in the fRC regime. 


\section{Comparison with spectroscopy}\label{sec:spec}

\begin{figure}
\centering
   \includegraphics[width=0.47\textwidth]{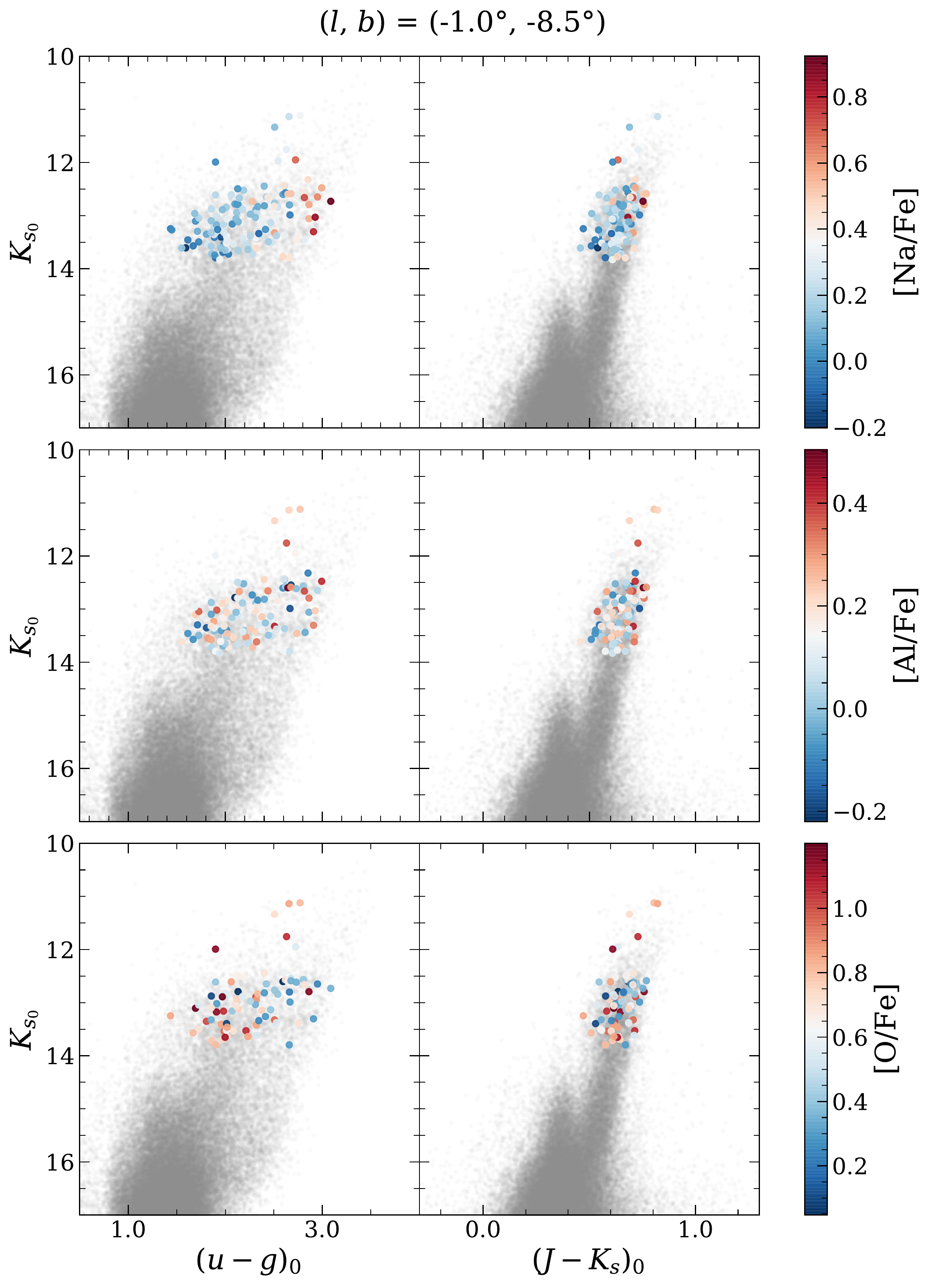}
     \caption{Same as Figure~\ref{fig:spec}, but for [Na/Fe], [Al/Fe], and [O/Fe] abundances in the ($K_{s}$, $u-g$)$_{0}$ and ($K_{s}$, $J-K_{s}$)$_{0}$ planes.
     The [Na/Fe] abundances of stars increase with increasing ($u-g$)$_{0}$ color, while the [O/Fe] abundances decrease. 
     These abundance gradients are not observed in the ($J-K_{s}$)$_{0}$ color. 
     In the case of [Al/Fe], a variation of abundance is not obvious in either ($u-g$)$_{0}$ or ($J-K_{s}$)$_{0}$. 
     }
     \label{fig:spec_other}
\end{figure}

\begin{figure}
\centering
   \includegraphics[width=0.47\textwidth]{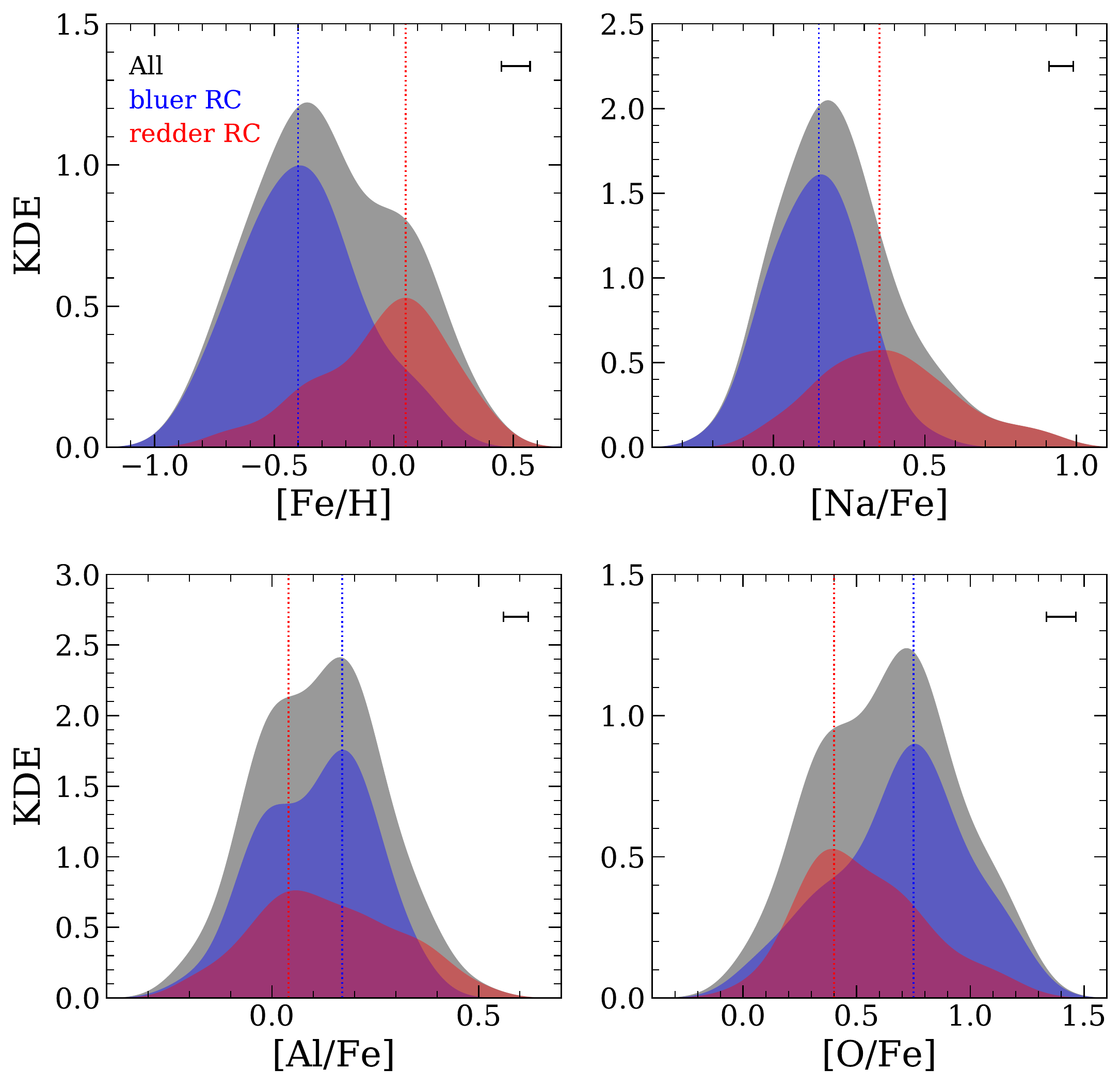}
     \caption{Kernel density estimates of [Fe/H], [Na/Fe], [Al/Fe], and [O/Fe] abundances for the bluer and redder stars.
     The redder stars are generally more enhanced in [Fe/H] and [Na/Fe] than the bluer stars, while this trend is reversed in [Al/Fe] and [O/Fe]. 
     The vertical red and blue dotted lines indicate the peak abundances of each group, and the bandwidth for KDE is in the upper right corner of each panel.
     }   
     \label{fig:spec_hist}
\end{figure}

In order to further examine the relationship between the colors of the RC stars and their chemical composition, we compare the BDBS data with high-resolution spectroscopic data. The spectroscopic data are from \citet{Lim2021}, and were obtained using the Michigan/Magellan Fiber System \citep[M2FS;][]{Mateo2012} on the Magellan telescope for the RC and RGB stars in the field of ($l$, $b$) = (-1$\degree$, -8.5$\degree$). For this comparison, we performed the same sample-selection procedure as that described in Section~\ref{sec:data}, this time for stars in this field from the BDBS, VVV, and Gaia data. A total of 124 stars were cross-matched with the spectroscopic data, and the metallicities of these stars are over-plotted on the CMDs in Figure~\ref{fig:spec}. As is expected, the [Fe/H] abundances of stars are gradually enhanced from -0.9 dex to +0.4 dex with increasing color index (i.e., from blue to red). This metallicity gradient with stellar color is particularly evident in ($u-g$)$_{0}$, ($g-r$)$_{0}$ , and ($u-i$)$_{0}$, but less clear in ($r-i$)$_{0}$ and ($J-K_{s}$)$_{0}$. We also estimate the color--metallicity relations for stars in the RC range (12.0 $\le$ $K_{s_{0}}$ $\le$ 14.0) in the upper panels of Figure~\ref{fig:spec}. The mean offsets between the observed and fitted [Fe/H] values are 0.14, 0.16, 0.22, 0.15, and 0.23 dex in ($u-g$)$_{0}$, ($g-r$)$_{0}$, ($r-i$)$_{0}$, ($u-i$)$_{0}$, and ($J-K_{s}$)$_{0}$ colors, respectively, and their standard deviations are 0.18, 0.20, 0.26, 0.19, and 0.28. The [Fe/H] of stars show tight correlations in the ($u-g$)$_{0}$, ($g-r$)$_{0}$, and ($u-i$)$_{0}$ colors with a low standard deviation ($\sigma$ $\le$ 0.2), while these correlations are less distinct in the ($r-i$)$_{0}$ and ($J-K_{s}$)$_{0}$ colors. The color--metallicity relations for the BDBS ($u-g$)$_{0}$, ($g-r$)$_{0}$, and ($u-i$)$_{0}$ colors are as follows:
\[[\mathrm{Fe/H}] =  (0.633\pm0.048) \times (u-g)_{0} - (1.665\pm0.105),\]
\[[\mathrm{Fe/H}] =  (2.068\pm0.191) \times (g-r)_{0} - (1.738\pm0.140),\]
\[[\mathrm{Fe/H}] =  (0.451\pm0.037) \times (u-i)_{0} - (1.726\pm0.118).\]
In particular, the relation with ($u-i$)$_{0}$ color is comparable to that determined by \citet{Johnson2020}, which reads [Fe/H] = 0.563($u-i$)$_{0}$ $-$ 2.074 (magenta dashed line in Figure~\ref{fig:spec}). Therefore, we can confirm that the bluer and redder stars, divided in ($u-g$)$_{0}$ in Section~\ref{sec:color}, have different mean metallicities, separating into  metal-poor and metal-rich populations of the bulge. However, the trend of metallicity with magnitude is not noticeable\footnote{It appears that the difference in metallicity between stars in the bright and faint RCs reported by \citet{Lim2021} is probably due to the dominance of the redder stars (metal-rich) in the bRC regime and that of the bluer stars (metal-poor) in the fRC regime (see also Section~\ref{sec:color}).}. 

Figure~\ref{fig:spec_other} shows the comparison of the BDBS photometry with chemical abundances of Na, Al, and O. The metal-rich redder stars are more enhanced in [Na/Fe] than the metal-poor bluer stars, while this trend is reversed in [O/Fe] in the ($K_{s}$, $u-g$)$_{0}$ CMD. However, these abundance gradients are not observed in the ($K_{s}$, $J-K_{s}$)$_{0}$ CMD as they for the [Fe/H] abundance in Figure~\ref{fig:spec}. The Na enhancement with a decline in O in the metal-rich stars is consistent with the typical chemical trends of the bulge stars reported by several studies; for example the [Na/Fe] is increased and [O/Fe] is decreased with increasing [Fe/H] \citep[e.g.,][]{Johnson2014, Zasowski2019}. On the other hand, the NUV photometry has been efficiently used to trace the multiple stellar populations with different chemical abundance in light elements, such as N and Na, for GCs \citep[e.g.,][]{Cummings2014, Savino2018}. In this regard, the clear gradients of Na and O abundances with ($u-g$)$_{0}$ color may reflect abundance variations in light elements together with the primary effect of different metallicity. Therefore, this result further implies that the BDBS data are also useful for studying multiple populations in GCs through specific color combinations, such as the $c_{y}$ index \cite[e.g.,][see also Figure~\ref{app:fig:ugi}]{Savino2018}. However, for the case of [Al/Fe],  the separation between the bluer and redder stars and the gradient of abundance is not clear, although the [Al/Fe] abundances of the bulge stars slightly decrease with increasing [Fe/H], as in [O/Fe]. This could be because the variation of Al abundance with metallicity is not large compared to the Na and O \citep[see Figure~2 of][]{Zasowski2019}. 

We also plot the kernel density estimates (KDEs) of [Fe/H], [Na/Fe], [Al/Fe], and [O/Fe] abundances for stars in the bluer and redder RC subgroups, respectively, in Figure~\ref{fig:spec_hist}. While all the cross-matched stars show a distinct bimodal distribution in [Fe/H], the bluer and redder stars are clearly separated with a peak difference of $\sim$0.45 dex (peak value of -0.4 dex for bluer RC stars; +0.05 dex for redder RC stars). This difference is comparable to that between the metal-poor and metal-rich components of the bulge reported by previous studies. Here we note that \citet{Ness2013} reported a mean [Fe/H] of -0.25 dex for the metal-poor component and +0.15 dex for the metal-rich component, and -0.4 dex and +0.3 dex of the peak values of [Fe/H] are presented for the metal-poor and metal-rich components by \citet{Zoccali2017}. This similarity supports the idea that the bluer and redder stars of this study correspond to the metal-poor and metal-rich components of the bulge, respectively. 

In addition, the metal-rich, redder stars are also generally more enhanced in [Na/Fe] but depleted in [Al/Fe] and [O/Fe] than the metal-poor bluer stars, although this trend is less clear in [Al/Fe]. The differences between the two groups are $\sim$0.2 dex in [Na/Fe], $\sim$0.13 dex in [Al/Fe], and $\sim$0.35 dex in [O/Fe]. These chemical characteristics are analogous to that observed in  GCs between the earlier and later stellar populations in terms of the Na-O anti-correlation \citep[see, e.g.,][]{Carretta2009, Bastian2018}. However, we note that the Na-enhancement with O-depletion of the metal-rich stars could be naturally expected in the bulge \citep[see, e.g.,][]{Johnson2014, Zasowski2019}. Although the bluer and redder stars in the bulge show a large difference in metallicity, unlike typical GCs, the intrinsic metallicity variations are also observed in some peculiar GCs, such as $\omega$-Centauri, Gaia~1, and Terzan~5 \citep{Johnson2010,Origlia2011,Massari2014,Mucciarelli2017,Simpson2017,Schiavon2017,Koch2018Gaia1}. In this regard, further research is necessary to determine whether these chemical properties simply reflect the general trends of the bulge stars or are associated with the multiple populations of GCs. 


\section{Discussion} \label{sec:discussion}

Here, we use  the BDBS data to show that the bright and faint RCs observed in the bulge have contrasting distributions in ($u-g$)$_{0}$ and ($u-i$)$_{0}$ colors with significant variations depending on Galactic longitude and latitude. In particular, the stars on the RC  could be efficiently divided into  bluer and redder stars according to their ($u-g$)$_{0}$ colors. The redder stars are characteristic of the double RC, in that they show constant magnitudes in the bright and faint clumps and a variation in number ratio with longitude, while the bluer stars are mainly placed in the fainter RC regime regardless of the studied field. We also confirm that the redder stars are more enhanced in [Fe/H] and [Na/Fe], but more depleted in [O/Fe] than the bluer stars through a comparison with our spectroscopy. Our result is consistent with previous studies showing that the MW bulge hosts a spheroidal shape comprising a metal-poor component and a boxy/peanut-shaped metal-rich component \citep{Rojas-Arriagada2014, Kunder2016}. 

\subsection{Effect of the bar}
Although our findings are reasonable in the context of the current understanding of the bulge, which comprises a metal-poor and a  metal-rich component, two other possible explanations should be considered. The first is that the redder RC originated from the metal-rich population of the Galactic bar. \citet{McWilliam2010} suggested that the influence of the bar component is insufficient to explain the constant magnitude of the bright and faint RCs in the various fields of the bulge. However, if we suppose the single clump of the redder stars in the fields of $l$ = +4.5$\degree$ and -4.5$\degree$, the clump in the positive longitude is 0.7 mag brighter than that in the negative longitude field (see Figures~\ref{fig:density} and \ref{fig:contour_ug}). This difference in magnitude is comparable to that expected for the 9$\degree$ separation in longitude of the tilted bar (see Section~5 of \citealt{McWilliam2010}). Thus, while the RC of the metal-poor bulge stars is placed on the bluer and fainter regime, the RC of the metal-rich bar stars is moved in the redder regime, which could explain the observed magnitude and color distribution, as well as their dependence on the Galactic position. However, the effect of the bar alone cannot explain the relatively distinct double RC of the redder stars in the fields at $l$ = 0$\degree$, where its bright clump shows a magnitude that is similar to the clump in the $l$ = +4.5$\degree$ fields. Moreover, the faint clump shows  magnitudes that are consistent with the clump in the $l$ = -4.5$\degree$ fields. If the redder stars were to originate from the bar, they would be expected to form a single clump (at $l$ = 0$\degree$) in between the magnitude range of the clump of the $l$ = +4.5$\degree$ and $l$ = -4.5$\degree$ fields. One other interesting feature is the magnitude variation of the bluer clump in the fields of $b$ = $-$6.0$\degree$. As shown in Figures~\ref{fig:contour_ug}, \ref{fig:contour_jk}, and \ref{fig:contour_ui}, the peak magnitude of the bluer clump increases with decreasing longitude from 13.0 ($l$ = +4.5$\degree$) to 13.5 ($l$ = $-$4.5$\degree$) $K_{s_{0}}$~mag, while no clear variation is observed in the $b$ = $-$7.5$\degree$ and $-$9.0$\degree$ fields. \citet{Kunder2020} reported that one population of metal-poor RR Lyrae stars traces the barred structure. In this regard, this feature may reflect the tilted bar structure among metal-poor stars, because the barred signature would be negligible at  higher latitude  ($b$ = $-$7.5$\degree$ and $-$9.0$\degree$).
 
\begin{figure}
\centering
   \includegraphics[width=0.47\textwidth]{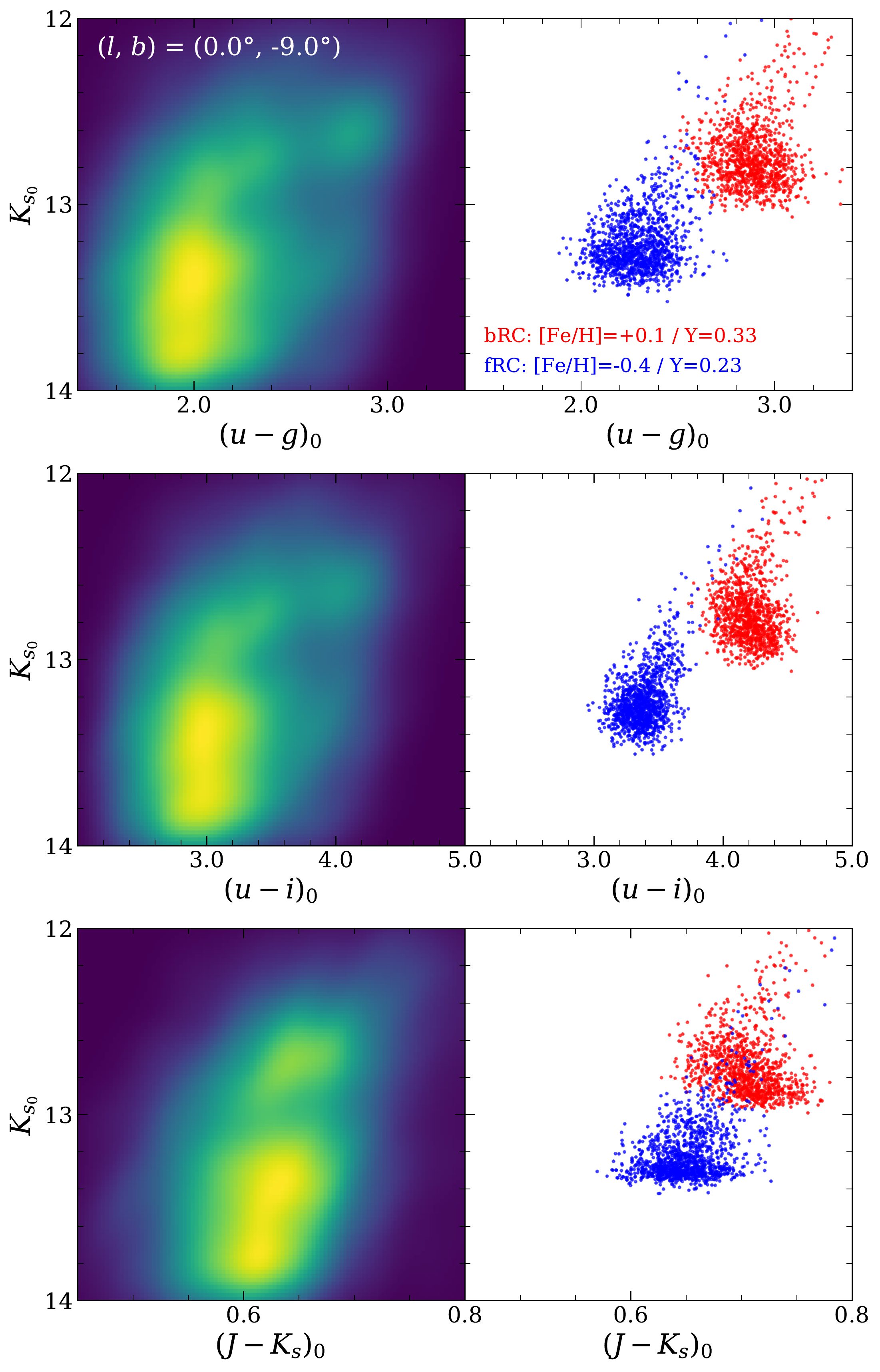}
     \caption{Comparison of our observation (left panels) with  synthetic population models (right panels).
     The difference in the ($u-g$)$_{0}$ and ($u-i$)$_{0}$ colors, with similar ($J-K_{s}$)$_{0}$ color, between the bRC and fRC can be reproduced with two-population models. 
     The bRC models (red circles) are enhanced not only in [Fe/H] but also in He abundance with respect to the fRC models (blue circles) with $\Delta$[FeH] = 0.5 dex and $\Delta$$Y$ = 0.1.
     }
     \label{fig:model}
\end{figure}

\subsection{Multiple-population scenario}
Another possible explanation for the observed color distribution could be the difference in chemical composition between the bright and faint RCs based on the multiple populations \citep[see][]{Lee2015, Joo2017, Lim2021}. The color of the RC is mainly related to metallicity, while the helium abundance of stars affects their luminosity. In this regard, the redder color of the bRC and the bluer color of the fRC could be well explained by the difference in both metallicity and He abundance of RC stars. Figure~\ref{fig:model} shows a comparison between observations of stars in the ($l$, $b$) = (0.0$\degree$, -9.0$\degree$) field with synthetic models for the two stellar populations with different chemical composition, but the same age of 10~Gyr and distance modulus of 15.0~mag in the $K_{s}$-band. These models are based on the evolutionary population synthesis of \citet{Chung2017}. As shown in this figure, the metal-poor and He-normal population of the fRC and the metal-rich and He-enhanced population of the bRC ($\Delta$[Fe/H] = 0.6 dex; $\Delta$ $Y$ = 0.1) nicely reproduce the observed features, such as the difference in the ($u-g$)$_{0}$ and ($u-i$)$_{0}$ with similar ($J-K_{s}$)$_{0}$ between the two RCs. The enhancement of He abundance in the later-generation stars has been reported in some massive GCs \citep[e.g.,][]{King2012, Milone2015}. In this scenario, the variation of the redder RC depending on longitude can be explained by the different influence of the bar component \citep[see][]{Joo2017}. Thus, the young and metal-rich bar component comprises the faint clump of the redder stars in the field of $l$ = 0.0$\degree$, while this component is embedded in the bright clump of the redder stars in the positive longitude fields, and forms the faint clump of the redder RC regime in the negative longitude fields. However, the lack of the redder bright clump in the fields of $l$ = -4.5$\degree$ and the almost identical magnitude of the redder faint clump between the $l$ = 0.0$\degree$ and -4.5$\degree$ fields remain to be explained. 

\subsection{X-shaped structure of the metal-rich component} 
In summary, our main findings for the color and luminosity distributions of the RC stars are as follows: 
(1) RC stars can efficiently be divided into metal-poor bluer and metal-rich redder stars in the ($u-g$)$_{0}$ color.  
(2) The metal-poor bluer stars populate a single RC with consistent magnitude in all fields. 
(3) The metal-rich redder stars show distinct double clumps at the fields of $l$ = 0.0$\degree$. 
(4) In these fields, the separation between the bright and faint clumps of the metal-rich redder stars is extended with increasing latitude,
and (5) the bright clump of the metal-rich redder stars becomes significant in the positive longitude fields, whereas the faint clump of the redder stars is dominant in the negative longitude fields at the same magnitude with the fields of $l$ = 0.0$\degree$. 
All these properties are best explained by a spheroidal shape of the metal-poor component and a boxy/peanut shape (X-shape) of the metal-rich component in the bulge. 

However, based on this scenario there is an additional issue to be addressed. As clearly shown in the field at ($l$, $b$) = (0.0$\degree$, -7.5$\degree$), the RC of the metal-poor stars is even fainter than the faint clump of the metal-rich stars (see Figure~\ref{fig:contour_ug}). If the double RC of the metal-rich stars originates from a different distance on the X-shaped structure, the spheroidal shape of the metal-poor stars should be placed in between the near and far arms of the X-structure. In this case, the fainter luminosity of the metal-poor component cannot be explained by a distance effect alone. This difference in luminosity could be due to metallicity or population effects on the RC. The RC stars become redder with increasing metallicity, while they become slightly brighter in NIR bands and fainter in optical bands \citep[see][]{Salaris2002}. This effect on luminosity increases with increasing age. For instance, when we compare  two stellar isochrones with the same age of 10~Gyr from the  ``Bag of Stellar Tracks and Isochrones'' \citep[BaSTI;][]{Hidalgo2018}, contrasting metal-poor ([M/H] = -0.401, $Z$ = 0.006, and $Y$ = 0.255) and metal-rich ([M/H]=+0.06, $Z$ = 0.017, and $Y$ = 0.269) cases, the RC of the metal-rich model is approximately 0.1 mag brighter in the K$_{s}$ band and 0.5 redder in ($u-g$) than that of the metal-poor model. We note that this difference in magnitude is comparable to that suggested by \setcitestyle{notesep={; }}\citet[][see their Table~1]{Salaris2002}. However, the metallicity effect seems to be insufficient to explain our observed magnitude difference between the bluer RC and the center of the redder double RC ($\sim$0.5 mag at ($l$, $b$) = (0.0$\degree$, -7.5$\degree$)). On the other hand, as the luminosity of RC stars is primarily affected by He abundance, a He-enhancement of the metal-rich component of the bulge may be required \citep[see][]{Joo2017}. Therefore, further study using population synthesis modeling is essential to examine whether the observed luminosity distribution of the metal-poor and metal-rich components can be reproduced with metallicity and distance differences only or whether an additional He abundance variation between the two components is required.
\\

The BDBS data allow an enormous sample of the bulge stars to be probed in six passbands ($ugrizY$), providing a new view of the luminosity and color distribution of the RC stars in the MW bulge. In particular, these data will be of great help in investigating the bimodal structure of the MW bulge composed of the metal-poor and metal-rich stellar populations. Further investigation of the detailed 3D structure model of the metal-poor and metal-rich populations of the bulge is required, and the BDBS is ideal for such studies.

 \vspace{5mm}
 \begin{acknowledgements}
We are grateful to the anonymous referee for a number of helpful suggestions and comments. 
DL and AJKH gratefully acknowledge funding by the Deutsche Forschungsgemeinschaft (DFG, German Research Foundation) -- SFB 881 (``The Milky Way System'', subprojects A03, A05, A11).
DL thanks Sree Oh for comments and encouragements.
This research made use of the cross-match service provided by CDS, Strasbourg.
Data used in this paper comes from the Blanco DECam Survey Collaboration. This project used data obtained with the Dark Energy Camera (DECam), which was constructed by the Dark Energy Survey (DES) collaboration. Funding for the DES Projects has been provided by the U.S. Department of Energy, the U.S. National Science Foundation, the Ministry of Science and Education of Spain, the Science and Technology Facilities Council of the United Kingdom, the Higher Education Funding Council for England, the National Center for Supercomputing Applications at the University of Illinois at Urbana-Champaign, the Kavli Institute of Cosmological Physics at the University of Chicago, the Center for Cosmology and Astro-Particle Physics at the Ohio State University, the Mitchell Institute for Fundamental Physics and Astronomy at Texas A\&M University, Financiadora de Estudos e Projetos, Funda\c{c}\~{o} Carlos Chagas Filho de Amparo \'a Pesquisa do Estado do Rio de Janeiro, Conselho Nacional de Desenvolvimento Cient\'{\i}fico e Tecnol\'ogico and the Minist\'erio da Ci\^encia, Tecnologia e Inovac\~{a}o, the Deutsche Forschungsgemeinschaft, and the Collaborating Institutions in the Dark Energy Survey. The Collaborating Institutions are Argonne National Laboratory, the University of California at Santa Cruz, the University of Cambridge, Centro de Investigaciones En\'ergeticas, Medioambientales y Tecnol\'ogicas-Madrid, the University of Chicago, University College London, the DES-Brazil Consortium, the University of Edinburgh, the Eidgen\"ossische Technische Hochschule (ETH) Z\"urich, Fermi National Accelerator Laboratory, the University of Illinois at Urbana-Champaign, the Institut de Ci\'oncies de l'Espai (IEEC/CSIC), the Institut de F\'{\i}sica d'Altes Energies, Lawrence Berkeley National Laboratory, the Ludwig-Maximilians Universit\"at M\"unchen and the associated Excellence Cluster Universe, the University of Michigan, the National Optical Astronomy Observatory, the University of Nottingham, the Ohio State University, the OzDES Membership Consortium the University of Pennsylvania, the University of Portsmouth, SLAC National Accelerator Laboratory, Stanford University, the University of Sussex, and Texas A\&M University. Based on observations at Cerro Tololo Inter-American Observatory (2013A-0529; 2014A-0480; PI: Rich), National Optical Astronomy Observatory, which is operated by the Association of Universities for Research in Astronomy (AURA) under a cooperative agreement with the National Science Foundation.
 \end{acknowledgements}


\bibliographystyle{aa} 
\bibliography{export-bibtex} 

\begin{thebibliography}{59}
\expandafter\ifx\csname natexlab\endcsname\relax\def\natexlab#1{#1}\fi

\bibitem[{{Alonso-Garc{\'\i}a} {et~al.}(2017){Alonso-Garc{\'\i}a}, {Minniti},
  {Catelan}, {Contreras Ramos}, {Gonzalez}, {Hempel}, {Lucas}, {Saito},
  {Valenti}, \& {Zoccali}}]{Alonso-Garcia2017}
{Alonso-Garc{\'\i}a}, J., {Minniti}, D., {Catelan}, M., {et~al.} 2017, \apjl,
  849, L13

\bibitem[{{Athanassoula}(2005)}]{Athanassoula2005}
{Athanassoula}, E. 2005, \mnras, 358, 1477

\bibitem[{{Babusiaux} {et~al.}(2010){Babusiaux}, {G{\'o}mez}, {Hill}, {Royer},
  {Zoccali}, {Arenou}, {Fux}, {Lecureur}, {Schultheis}, {Barbuy}, {Minniti}, \&
  {Ortolani}}]{Babusiaux2010}
{Babusiaux}, C., {G{\'o}mez}, A., {Hill}, V., {et~al.} 2010, \aap, 519, A77

\bibitem[{{Bastian} \& {Lardo}(2018)}]{Bastian2018}
{Bastian}, N. \& {Lardo}, C. 2018, \araa, 56, 83

\bibitem[{{Bureau} {et~al.}(2006){Bureau}, {Aronica}, {Athanassoula},
  {Dettmar}, {Bosma}, \& {Freeman}}]{Bureau2006}
{Bureau}, M., {Aronica}, G., {Athanassoula}, E., {et~al.} 2006, \mnras, 370,
  753

\bibitem[{{Buta} {et~al.}(2015){Buta}, {Sheth}, {Athanassoula}, {Bosma},
  {Knapen}, {Laurikainen}, {Salo}, {Elmegreen}, {Ho}, {Zaritsky}, {Courtois},
  {Hinz}, {Mu{\~n}oz-Mateos}, {Kim}, {Regan}, {Gadotti}, {Gil de Paz}, {Laine},
  {Men{\'e}ndez-Delmestre}, {Comer{\'o}n}, {Erroz Ferrer}, {Seibert},
  {Mizusawa}, {Holwerda}, \& {Madore}}]{Buta2015}
{Buta}, R.~J., {Sheth}, K., {Athanassoula}, E., {et~al.} 2015, \apjs, 217, 32

\bibitem[{{Carretta} {et~al.}(2009){Carretta}, {Bragaglia}, {Gratton},
  {Lucatello}, {Catanzaro}, {Leone}, {Bellazzini}, {Claudi}, {D'Orazi},
  {Momany}, {Ortolani}, {Pancino}, {Piotto}, {Recio-Blanco}, \&
  {Sabbi}}]{Carretta2009}
{Carretta}, E., {Bragaglia}, A., {Gratton}, R.~G., {et~al.} 2009, \aap, 505,
  117

\bibitem[{{Casagrande} \& {VandenBerg}(2018)}]{Casagrande2018}
{Casagrande}, L. \& {VandenBerg}, D.~A. 2018, \mnras, 479, L102

\bibitem[{{Chung} {et~al.}(2017){Chung}, {Yoon}, \& {Lee}}]{Chung2017}
{Chung}, C., {Yoon}, S.-J., \& {Lee}, Y.-W. 2017, \apj, 842, 91

\bibitem[{{Clarkson} {et~al.}(2018){Clarkson}, {Calamida}, {Sahu}, {Brown},
  {Gennaro}, {Avila}, {Valenti}, {Debattista}, {Rich}, {Minniti}, {Zoccali}, \&
  {Aufdemberge}}]{Clarkson2018}
{Clarkson}, W.~I., {Calamida}, A., {Sahu}, K.~C., {et~al.} 2018, \apj, 858, 46

\bibitem[{{Cummings} {et~al.}(2014){Cummings}, {Geisler}, {Villanova}, \&
  {Carraro}}]{Cummings2014}
{Cummings}, J.~D., {Geisler}, D., {Villanova}, S., \& {Carraro}, G. 2014, \aj,
  148, 27

\bibitem[{{Gaia Collaboration} {et~al.}(2018){Gaia Collaboration}, {Brown},
  {Vallenari}, {Prusti}, {de Bruijne}, {Babusiaux}, {Bailer-Jones}, {Biermann},
  {Evans}, {Eyer}, \& et~al.}]{GaiaCollaboration2018}
{Gaia Collaboration}, {Brown}, A.~G.~A., {Vallenari}, A., {et~al.} 2018, \aap,
  616, A1

\bibitem[{{Gonzalez} {et~al.}(2016){Gonzalez}, {Gadotti}, {Debattista},
  {Rejkuba}, {Valenti}, {Zoccali}, {Coccato}, {Minniti}, \&
  {Ness}}]{Gonzalez2016}
{Gonzalez}, O.~A., {Gadotti}, D.~A., {Debattista}, V.~P., {et~al.} 2016, \aap,
  591, A7

\bibitem[{{Gratton} {et~al.}(2012){Gratton}, {Carretta}, \&
  {Bragaglia}}]{Gratton2012}
{Gratton}, R.~G., {Carretta}, E., \& {Bragaglia}, A. 2012, \aapr, 20, 50

\bibitem[{{Green} {et~al.}(2018){Green}, {Schlafly}, {Finkbeiner}, {Rix},
  {Martin}, {Burgett}, {Draper}, {Flewelling}, {Hodapp}, {Kaiser}, {Kudritzki},
  {Magnier}, {Metcalfe}, {Tonry}, {Wainscoat}, \& {Waters}}]{Green2018}
{Green}, G.~M., {Schlafly}, E.~F., {Finkbeiner}, D., {et~al.} 2018, \mnras,
  478, 651

\bibitem[{{Hidalgo} {et~al.}(2018){Hidalgo}, {Pietrinferni}, {Cassisi},
  {Salaris}, {Mucciarelli}, {Savino}, {Aparicio}, {Silva Aguirre}, \&
  {Verma}}]{Hidalgo2018}
{Hidalgo}, S.~L., {Pietrinferni}, A., {Cassisi}, S., {et~al.} 2018, \apj, 856,
  125

\bibitem[{{Johnson} \& {Pilachowski}(2010)}]{Johnson2010}
{Johnson}, C.~I. \& {Pilachowski}, C.~A. 2010, \apj, 722, 1373

\bibitem[{{Johnson} {et~al.}(2014){Johnson}, {Rich}, {Kobayashi}, {Kunder}, \&
  {Koch}}]{Johnson2014}
{Johnson}, C.~I., {Rich}, R.~M., {Kobayashi}, C., {Kunder}, A., \& {Koch}, A.
  2014, \aj, 148, 67

\bibitem[{{Johnson} {et~al.}(2020){Johnson}, {Rich}, {Young}, {Simion},
  {Clarkson}, {Pilachowski}, {Michael}, {Kunder}, {Koch}, \&
  {Vivas}}]{Johnson2020}
{Johnson}, C.~I., {Rich}, R.~M., {Young}, M.~D., {et~al.} 2020, \mnras, 499,
  2357

\bibitem[{{Joo} {et~al.}(2017){Joo}, {Lee}, \& {Chung}}]{Joo2017}
{Joo}, S.-J., {Lee}, Y.-W., \& {Chung}, C. 2017, \apj, 840, 98

\bibitem[{{King} {et~al.}(2012){King}, {Bedin}, {Cassisi}, {Milone}, {Bellini},
  {Piotto}, {Anderson}, {Pietrinferni}, \& {Cordier}}]{King2012}
{King}, I.~R., {Bedin}, L.~R., {Cassisi}, S., {et~al.} 2012, \aj, 144, 5

\bibitem[{{Koch} {et~al.}(2018){Koch}, {Hansen}, \& {Kunder}}]{Koch2018Gaia1}
{Koch}, A., {Hansen}, T.~T., \& {Kunder}, A. 2018, \aap, 609, A13

\bibitem[{{Koch} {et~al.}(2016){Koch}, {McWilliam}, {Preston}, \&
  {Thompson}}]{Koch2016}
{Koch}, A., {McWilliam}, A., {Preston}, G.~W., \& {Thompson}, I.~B. 2016, \aap,
  587, A124

\bibitem[{{Kunder} {et~al.}(2020){Kunder}, {P{\'e}rez-Villegas}, {Rich},
  {Ogata}, {Murari}, {Boren}, {Johnson}, {Nataf}, {Walker}, {Bono}, {Koch},
  {Propris}, {Storm}, \& {Wojno}}]{Kunder2020}
{Kunder}, A., {P{\'e}rez-Villegas}, A., {Rich}, R.~M., {et~al.} 2020, \aj, 159,
  270

\bibitem[{{Kunder} {et~al.}(2016){Kunder}, {Rich}, {Koch}, {Storm}, {Nataf},
  {De Propris}, {Walker}, {Bono}, {Johnson}, {Shen}, \& {Li}}]{Kunder2016}
{Kunder}, A., {Rich}, R.~M., {Koch}, A., {et~al.} 2016, \apjl, 821, L25

\bibitem[{{Lee} {et~al.}(2013){Lee}, {Han}, {Joo}, {Jang}, {Na}, {Okamoto},
  {Arimoto}, {Lim}, {Kim}, \& {Yoon}}]{Lee2013}
{Lee}, Y.-W., {Han}, S.-I., {Joo}, S.-J., {et~al.} 2013, \apjl, 778, L13

\bibitem[{{Lee} {et~al.}(2018){Lee}, {Hong}, {Lim}, {Chung}, {Jang}, {Kim}, \&
  {Joo}}]{Lee2018}
{Lee}, Y.-W., {Hong}, S., {Lim}, D., {et~al.} 2018, \apjl, 862, L8

\bibitem[{{Lee} {et~al.}(2015){Lee}, {Joo}, \& {Chung}}]{Lee2015}
{Lee}, Y.-W., {Joo}, S.-J., \& {Chung}, C. 2015, \mnras, 453, 3906

\bibitem[{{Lim} {et~al.}(2021){Lim}, {Lee}, {Koch}, {Hong}, {Johnson}, {Kim},
  {Chung}, {Mateo}, \& {Bailey}}]{Lim2021}
{Lim}, D., {Lee}, Y.-W., {Koch}, A., {et~al.} 2021, ApJ, in press,
  arXiv:2012.03954

\bibitem[{{Massari} {et~al.}(2014){Massari}, {Mucciarelli}, {Ferraro},
  {Origlia}, {Rich}, {Lanzoni}, {Dalessand ro}, {Valenti}, {Ibata}, {Lovisi},
  {Bellazzini}, \& {Reitzel}}]{Massari2014}
{Massari}, D., {Mucciarelli}, A., {Ferraro}, F.~R., {et~al.} 2014, \apj, 795,
  22

\bibitem[{{Mateo} {et~al.}(2012){Mateo}, {Bailey}, {Crane}, {Shectman},
  {Thompson}, {Roederer}, {Bigelow}, \& {Gunnels}}]{Mateo2012}
{Mateo}, M., {Bailey}, J.~I., {Crane}, J., {et~al.} 2012, in Society of
  Photo-Optical Instrumentation Engineers (SPIE) Conference Series, Vol. 8446,
  Ground-based and Airborne Instrumentation for Astronomy IV, 84464Y

\bibitem[{{McWilliam} \& {Zoccali}(2010)}]{McWilliam2010}
{McWilliam}, A. \& {Zoccali}, M. 2010, \apj, 724, 1491

\bibitem[{{Milone}(2015)}]{Milone2015}
{Milone}, A.~P. 2015, \mnras, 446, 1672

\bibitem[{{Minniti} {et~al.}(2010){Minniti}, {Lucas}, {Emerson}, {Saito},
  {Hempel}, {Pietrukowicz}, {Ahumada}, {Alonso}, {Alonso-Garcia}, {Arias},
  {Bandyopadhyay}, {Barb{\'a}}, {Barbuy}, {Bedin}, {Bica}, {Borissova},
  {Bronfman}, {Carraro}, {Catelan}, {Clari{\'a}}, {Cross}, {de Grijs},
  {D{\'e}k{\'a}ny}, {Drew}, {Fari{\~n}a}, {Feinstein}, {Fern{\'a}ndez
  Laj{\'u}s}, {Gamen}, {Geisler}, {Gieren}, {Goldman}, {Gonzalez}, {Gunthardt},
  {Gurovich}, {Hambly}, {Irwin}, {Ivanov}, {Jord{\'a}n}, {Kerins}, {Kinemuchi},
  {Kurtev}, {L{\'o}pez-Corredoira}, {Maccarone}, {Masetti}, {Merlo},
  {Messineo}, {Mirabel}, {Monaco}, {Morelli}, {Padilla}, {Palma}, {Parisi},
  {Pignata}, {Rejkuba}, {Roman-Lopes}, {Sale}, {Schreiber}, {Schr{\"o}der},
  {Smith}, {}, {Soto}, {Tamura}, {Tappert}, {Thompson}, {Toledo}, {Zoccali}, \&
  {Pietrzynski}}]{Minniti2010}
{Minniti}, D., {Lucas}, P.~W., {Emerson}, J.~P., {et~al.} 2010, \na, 15, 433

\bibitem[{{Mucciarelli} {et~al.}(2017){Mucciarelli}, {Monaco}, {Bonifacio}, \&
  {Saviane}}]{Mucciarelli2017}
{Mucciarelli}, A., {Monaco}, L., {Bonifacio}, P., \& {Saviane}, I. 2017, \aap,
  603, L7

\bibitem[{{Nataf}(2017)}]{Nataf2017}
{Nataf}, D.~M. 2017, \pasa, 34, e041

\bibitem[{{Nataf} {et~al.}(2014){Nataf}, {Cassisi}, \&
  {Athanassoula}}]{Nataf2014}
{Nataf}, D.~M., {Cassisi}, S., \& {Athanassoula}, E. 2014, \mnras, 442, 2075

\bibitem[{{Nataf} {et~al.}(2010){Nataf}, {Udalski}, {Gould}, {Fouqu{\'e}}, \&
  {Stanek}}]{Nataf2010}
{Nataf}, D.~M., {Udalski}, A., {Gould}, A., {Fouqu{\'e}}, P., \& {Stanek},
  K.~Z. 2010, \apjl, 721, L28

\bibitem[{{Nataf} {et~al.}(2015){Nataf}, {Udalski}, {Skowron}, {Szyma{\'n}ski},
  {Kubiak}, {Pietrzy{\'n}ski}, {Soszy{\'n}ski}, {Ulaczyk}, {Wyrzykowski},
  {Poleski}, {Athanassoula}, {Ness}, {Shen}, \& {Li}}]{Nataf2015}
{Nataf}, D.~M., {Udalski}, A., {Skowron}, J., {et~al.} 2015, \mnras, 447, 1535

\bibitem[{{Ness} {et~al.}(2013){Ness}, {Freeman}, {Athanassoula},
  {Wylie-de-Boer}, {Bland-Hawthorn}, {Asplund}, {Lewis}, {Yong}, {Lane}, \&
  {Kiss}}]{Ness2013}
{Ness}, M., {Freeman}, K., {Athanassoula}, E., {et~al.} 2013, \mnras, 430, 836

\bibitem[{{Ness} {et~al.}(2012){Ness}, {Freeman}, {Athanassoula},
  {Wylie-De-Boer}, {Bland-Hawthorn}, {Lewis}, {Yong}, {Asplund}, {Lane},
  {Kiss}, \& {Ibata}}]{Ness2012}
{Ness}, M., {Freeman}, K., {Athanassoula}, E., {et~al.} 2012, \apj, 756, 22

\bibitem[{{Origlia} {et~al.}(2011){Origlia}, {Rich}, {Ferraro}, {Lanzoni},
  {Bellazzini}, {Dalessandro}, {Mucciarelli}, {Valenti}, \&
  {Beccari}}]{Origlia2011}
{Origlia}, L., {Rich}, R.~M., {Ferraro}, F.~R., {et~al.} 2011, \apjl, 726, L20

\bibitem[{{Rattenbury} {et~al.}(2007){Rattenbury}, {Mao}, {Sumi}, \&
  {Smith}}]{Rattenbury2007}
{Rattenbury}, N.~J., {Mao}, S., {Sumi}, T., \& {Smith}, M.~C. 2007, \mnras,
  378, 1064

\bibitem[{{Rich} {et~al.}(2020){Rich}, {Johnson}, {Young}, {Simion},
  {Clarkson}, {Pilachowski}, {Michael}, {Kunder}, {Katherina Vivas}, {Koch},
  {Marchetti}, {Ibata}, {Martin}, {Robin}, {Lagarde}, {Collins}, {Ivezi{\'c}},
  {de Propris}, {Shen}, {Gerhard}, \& {Soto}}]{Rich2020}
{Rich}, R.~M., {Johnson}, C.~I., {Young}, M., {et~al.} 2020, \mnras, 499, 2340

\bibitem[{{Rojas-Arriagada} {et~al.}(2014){Rojas-Arriagada}, {Recio-Blanco},
  {Hill}, {de Laverny}, {Schultheis}, {Babusiaux}, {Zoccali}, {Minniti},
  {Gonzalez}, {Feltzing}, {Gilmore}, {Randich}, {Vallenari}, {Alfaro},
  {Bensby}, {Bragaglia}, {Flaccomio}, {Lanzafame}, {Pancino}, {Smiljanic},
  {Bergemann}, {Costado}, {Damiani}, {Hourihane}, {Jofr{\'e}}, {Lardo},
  {Magrini}, {Maiorca}, {Morbidelli}, {Sbordone}, {Worley}, {Zaggia}, \&
  {Wyse}}]{Rojas-Arriagada2014}
{Rojas-Arriagada}, A., {Recio-Blanco}, A., {Hill}, V., {et~al.} 2014, \aap,
  569, A103

\bibitem[{{Salaris} \& {Girardi}(2002)}]{Salaris2002}
{Salaris}, M. \& {Girardi}, L. 2002, \mnras, 337, 332

\bibitem[{{Savino} {et~al.}(2020){Savino}, {Koch}, {Prudil}, {Kunder}, \&
  {Smolec}}]{Savino2020}
{Savino}, A., {Koch}, A., {Prudil}, Z., {Kunder}, A., \& {Smolec}, R. 2020,
  \aap, 641, A96

\bibitem[{{Savino} {et~al.}(2018){Savino}, {Massari}, {Bragaglia}, {Dalessand
  ro}, \& {Tolstoy}}]{Savino2018}
{Savino}, A., {Massari}, D., {Bragaglia}, A., {Dalessand ro}, E., \& {Tolstoy},
  E. 2018, \mnras, 474, 4438

\bibitem[{{Schiavon} {et~al.}(2017){Schiavon}, {Johnson}, {Frinchaboy},
  {Zasowski}, {M{\'e}sz{\'a}ros}, {Garc{\'\i}a-Hern{\'a}ndez}, {Cohen}, {Tang},
  {Villanova}, {Geisler}, {Beers}, {Fern{\'a}ndez-Trincado}, {Garc{\'\i}a
  P{\'e}rez}, {Lucatello}, {Majewski}, {Martell}, {O'Connell}, {Allende
  Prieto}, {Bizyaev}, {Carrera}, {Lane}, {Malanushenko}, {Malanushenko},
  {Mu{\~n}oz}, {Nitschelm}, {Oravetz}, {Pan}, {Roman-Lopes}, {Schultheis}, \&
  {Simmons}}]{Schiavon2017}
{Schiavon}, R.~P., {Johnson}, J.~A., {Frinchaboy}, P.~M., {et~al.} 2017,
  \mnras, 466, 1010

\bibitem[{{Schlafly} \& {Finkbeiner}(2011)}]{Schlafly2011}
{Schlafly}, E.~F. \& {Finkbeiner}, D.~P. 2011, \apj, 737, 103

\bibitem[{{Simion} {et~al.}(2017){Simion}, {Belokurov}, {Irwin}, {Koposov},
  {Gonzalez-Fernandez}, {Robin}, {Shen}, \& {Li}}]{Simion2017}
{Simion}, I.~T., {Belokurov}, V., {Irwin}, M., {et~al.} 2017, \mnras, 471, 4323

\bibitem[{{Simpson} {et~al.}(2017){Simpson}, {De Silva}, {Martell}, {Zucker},
  {Ferguson}, {Bernard}, {Irwin}, {Penarrubia}, \& {Tolstoy}}]{Simpson2017}
{Simpson}, J.~D., {De Silva}, G.~M., {Martell}, S.~L., {et~al.} 2017, \mnras,
  471, 4087

\bibitem[{{Stanek} {et~al.}(1994){Stanek}, {Mateo}, {Udalski}, {Szymanski},
  {Kaluzny}, \& {Kubiak}}]{Stanek1994}
{Stanek}, K.~Z., {Mateo}, M., {Udalski}, A., {et~al.} 1994, \apjl, 429, L73

\bibitem[{{Taylor}(2005)}]{Taylor2005}
{Taylor}, M.~B. 2005, in Astronomical Society of the Pacific Conference Series,
  Vol. 347, Astronomical Data Analysis Software and Systems XIV, ed.
  P.~{Shopbell}, M.~{Britton}, \& R.~{Ebert}, 29

\bibitem[{{Uttenthaler} {et~al.}(2012){Uttenthaler}, {Schultheis}, {Nataf},
  {Robin}, {Lebzelter}, \& {Chen}}]{Uttenthaler2012}
{Uttenthaler}, S., {Schultheis}, M., {Nataf}, D.~M., {et~al.} 2012, \aap, 546,
  A57

\bibitem[{{Wegg} \& {Gerhard}(2013)}]{Wegg2013}
{Wegg}, C. \& {Gerhard}, O. 2013, \mnras, 435, 1874

\bibitem[{{Wegg} {et~al.}(2015){Wegg}, {Gerhard}, \& {Portail}}]{Wegg2015}
{Wegg}, C., {Gerhard}, O., \& {Portail}, M. 2015, \mnras, 450, 4050

\bibitem[{{Zasowski} {et~al.}(2019){Zasowski}, {Schultheis}, {Hasselquist},
  {Cunha}, {Sobeck}, {Johnson}, {Rojas-Arriagada}, {Majewski}, {Andrews},
  {J{\"o}nsson}, {Beers}, {Chojnowski}, {Frinchaboy}, {Holtzman}, {Minniti},
  {Nidever}, \& {Nitschelm}}]{Zasowski2019}
{Zasowski}, G., {Schultheis}, M., {Hasselquist}, S., {et~al.} 2019, \apj, 870,
  138

\bibitem[{{Zoccali} {et~al.}(2017){Zoccali}, {Vasquez}, {Gonzalez}, {Valenti},
  {Rojas-Arriagada}, {Minniti}, {Rejkuba}, {Minniti}, {McWilliam}, {Babusiaux},
  {Hill}, \& {Renzini}}]{Zoccali2017}
{Zoccali}, M., {Vasquez}, S., {Gonzalez}, O.~A., {et~al.} 2017, \aap, 599, A12

\end{thebibliography}

\begin{appendix}
\section{Density maps from various colors} \label{sec:appendix}
\begin{figure}
\centering
   \includegraphics[width=0.50\textwidth]{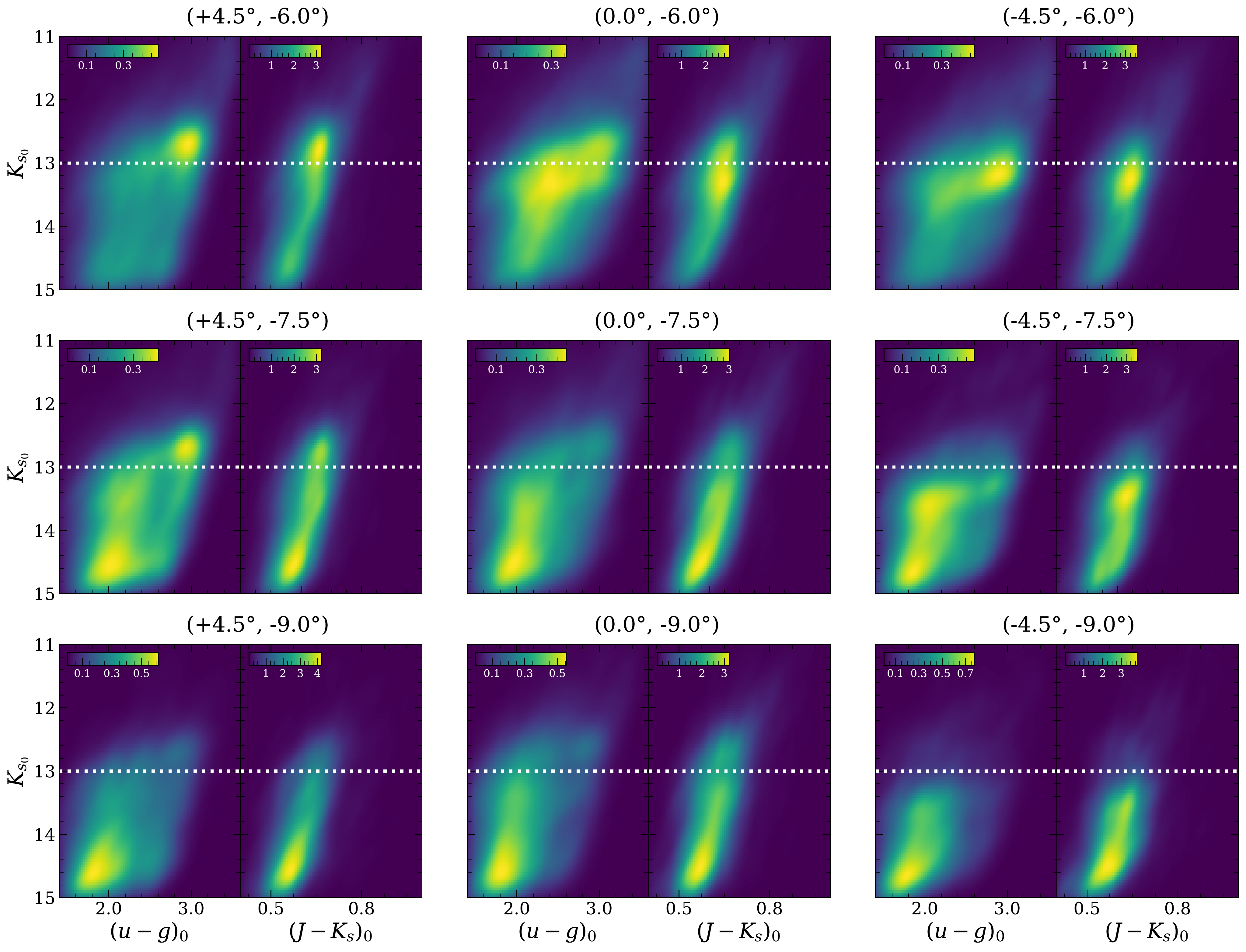}
   \caption{Density maps of stars in the RGB regime for nine Galactic fields in the ($u-g$)$_{0}$ and ($J-K_{s}$)$_{0}$ color planes.
   A clear split of RGBs is shown in the field of ($l$, $b$) = (+4.5$\degree$, -7.5$\degree$).}
   \label{app:fig:rgb}
\end{figure}

\begin{figure}
\centering
   \includegraphics[width=0.47\textwidth]{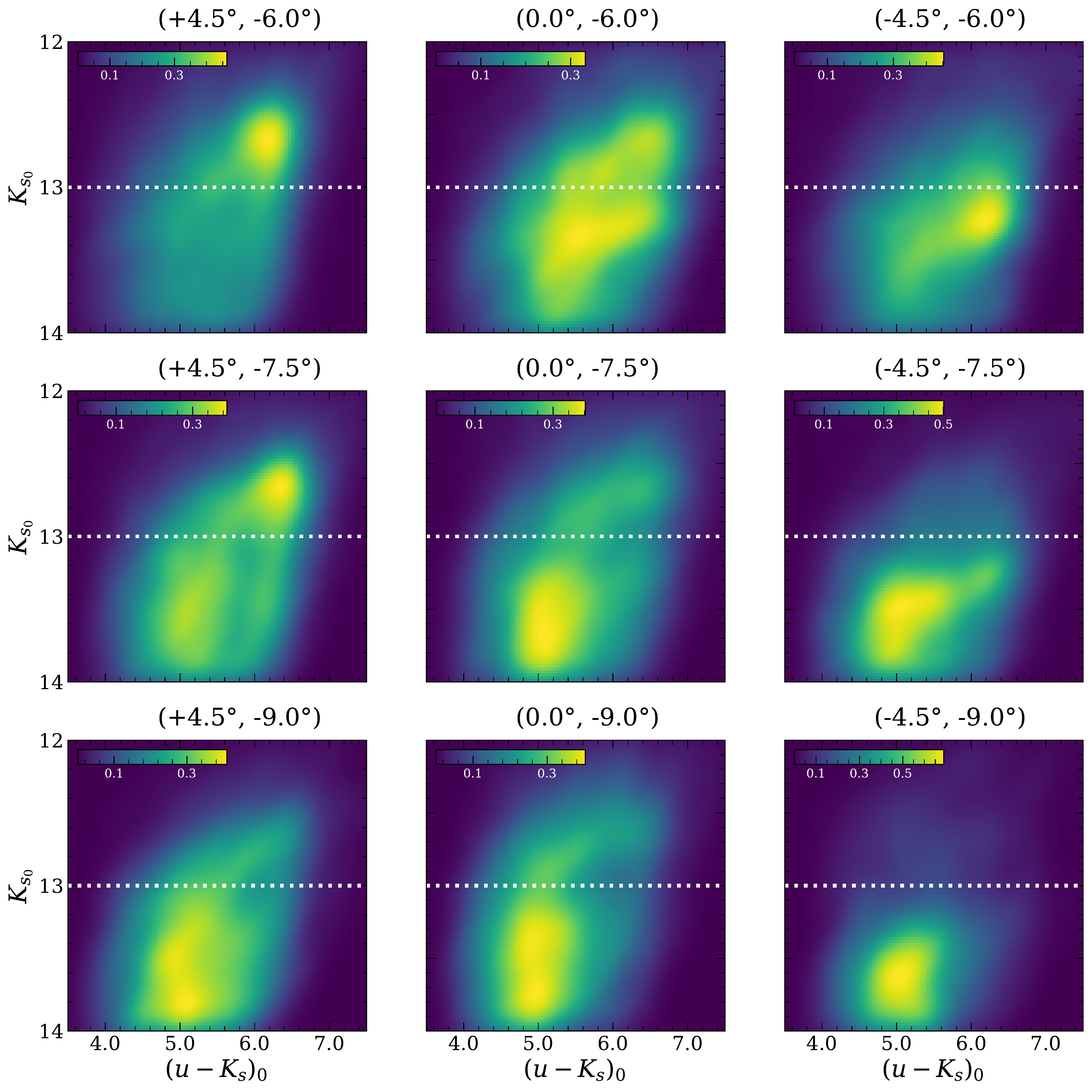}
   \caption{Density maps of stars in the RC regime for nine Galactic fields in the ($u-K_{s}$)$_{0}$ color plane.
   Similar to the cases of ($u-g$)$_{0}$ and ($u-i$)$_{0}$ colors, contrasting distributions between the bright and faint RC are shown.}
   \label{app:fig:uk}
\end{figure}

\begin{figure}
\centering
   \includegraphics[width=0.47\textwidth]{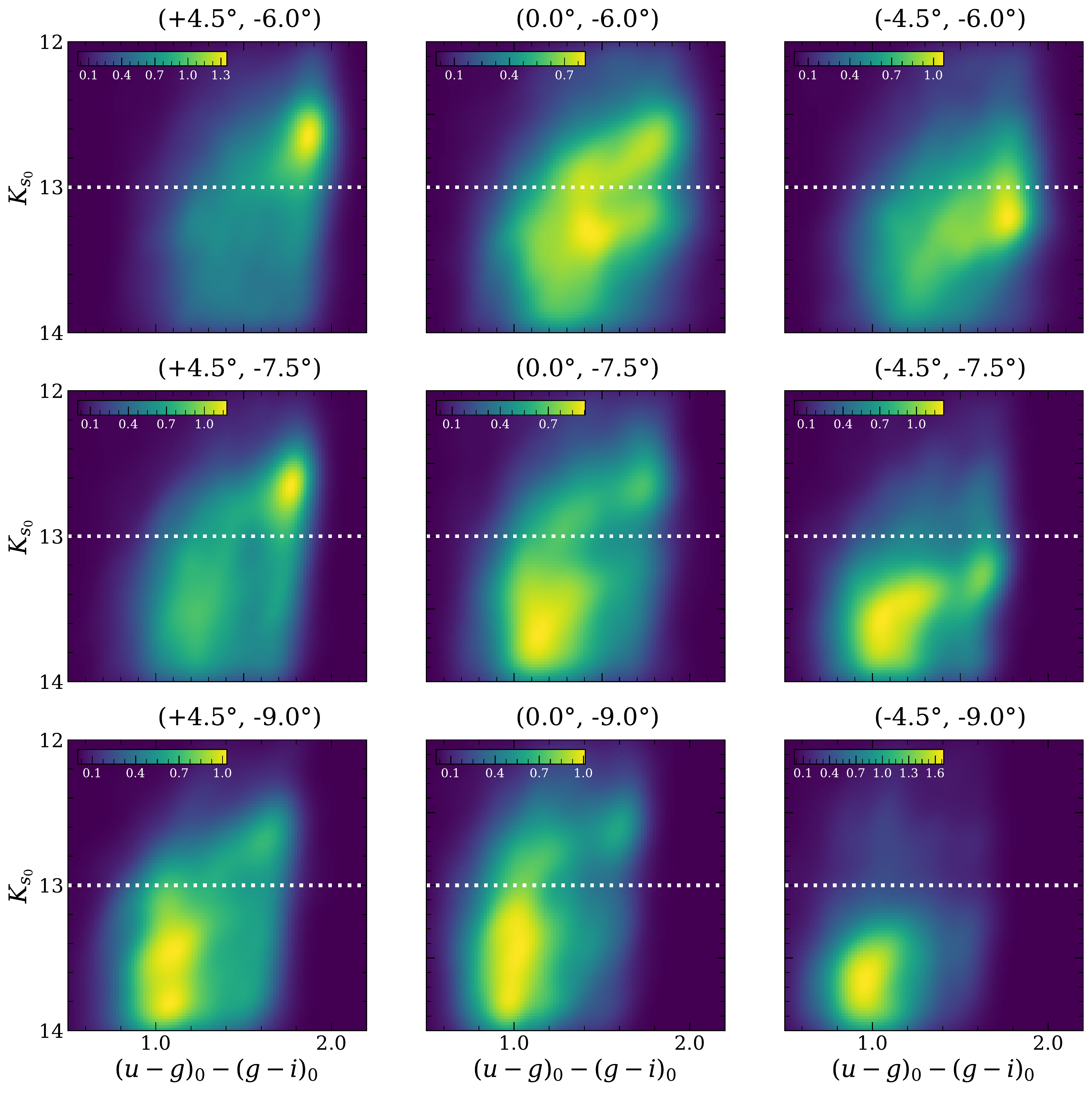}
   \caption{Same as Figure~\ref{app:fig:uk}, but in the ($u-g$)$_{0}$-($g-i$)$_{0}$ color plane.
   The overall patterns are consistent with those in ($u-g$)$_{0}$ and ($u-K_{s}$)$_{0}$ colors.  
   In particular, the separation between the blue and redder RC stars is more distinct in this color combination.}
   \label{app:fig:ugi}
\end{figure}

\begin{figure}
\centering
   \includegraphics[width=0.47\textwidth]{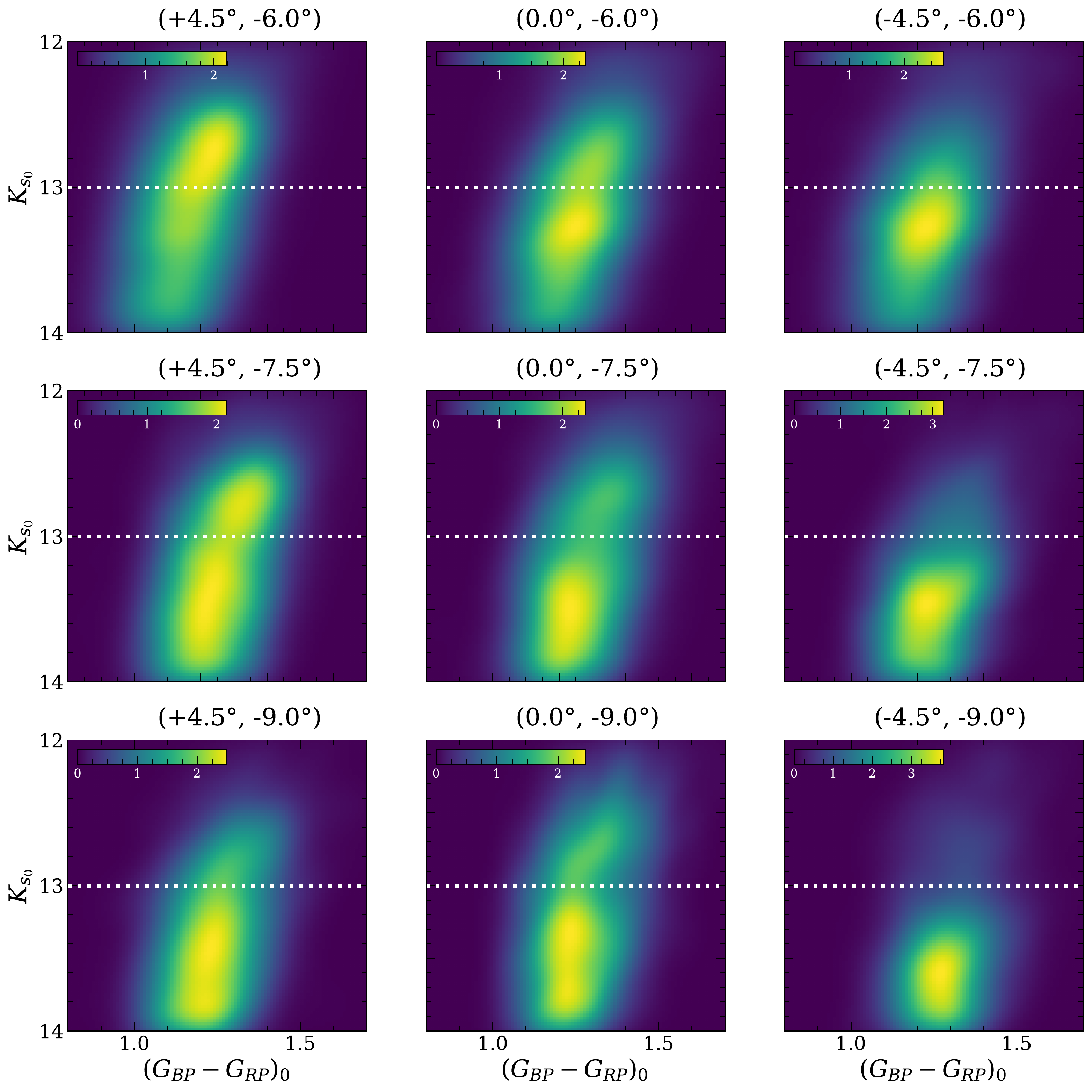}
   \caption{Same as Figure~\ref{app:fig:uk}, but in the ($G_{BP}-G_{Rp}$)$_{0}$ color plane.
   Although the double RC feature is shown, no difference in the color distribution between the bright and faint RC stars is observed.}
   \label{app:fig:br}
\end{figure}

\begin{figure}
\centering
   \includegraphics[width=0.47\textwidth]{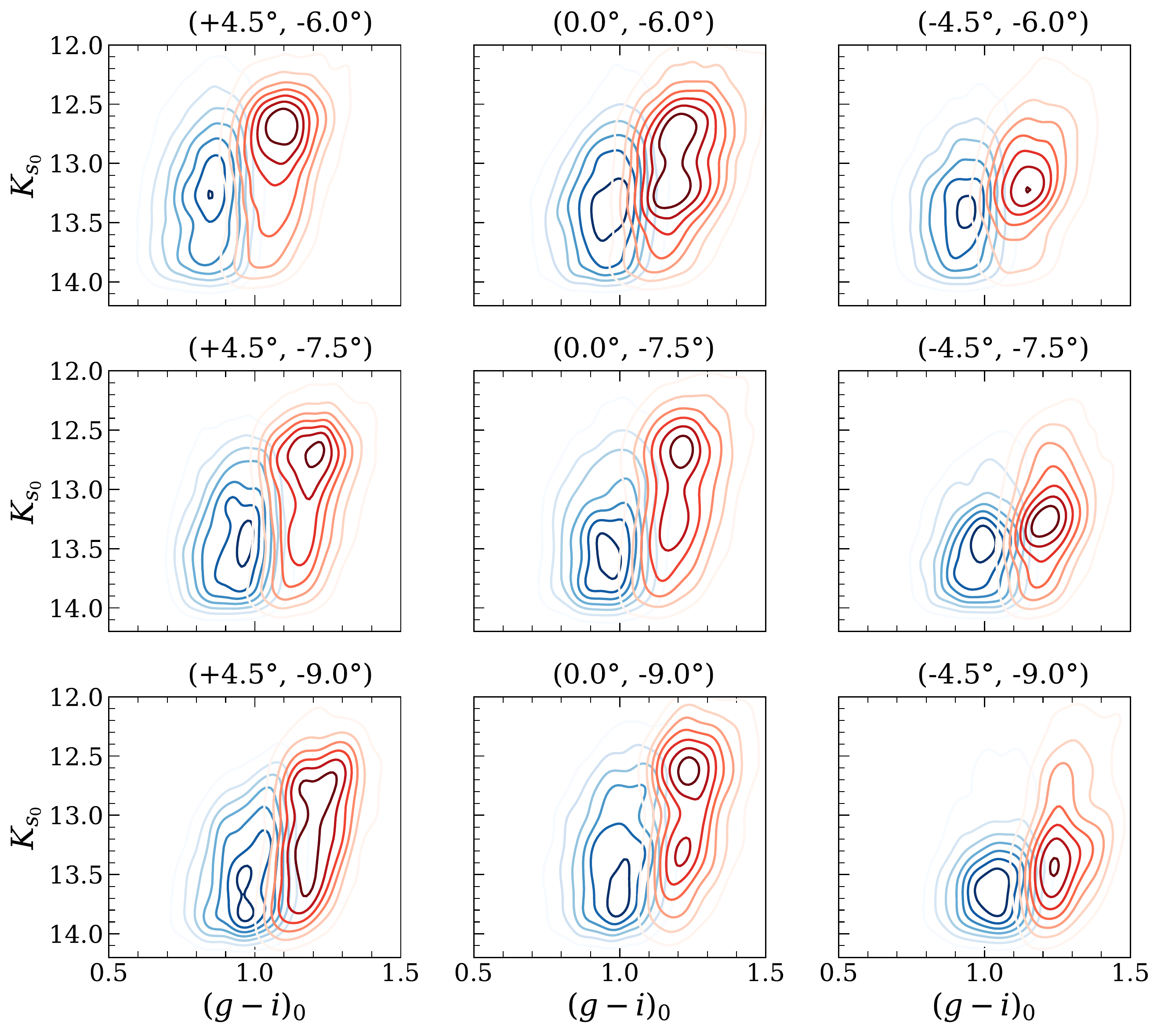}
   \caption{Density contours of stars in the RC regime with subgrouping by ($u-g$)$_{0}$ color, in the ($K_{s}$, $g-i$)$_{0}$ plane. 
   Similar to the ($u-g$)$_{0}$ and ($u-i$)$_{0}$ colors in Figures~\ref{fig:contour_ug} and \ref{fig:contour_ui}, bluer and redder RC stars are well divided in this color plane with a single clump of bluer stars and a double clump of redder stars.}
   \label{app:fig:gi}
\end{figure}

\begin{figure}
\centering
   \includegraphics[width=0.47\textwidth]{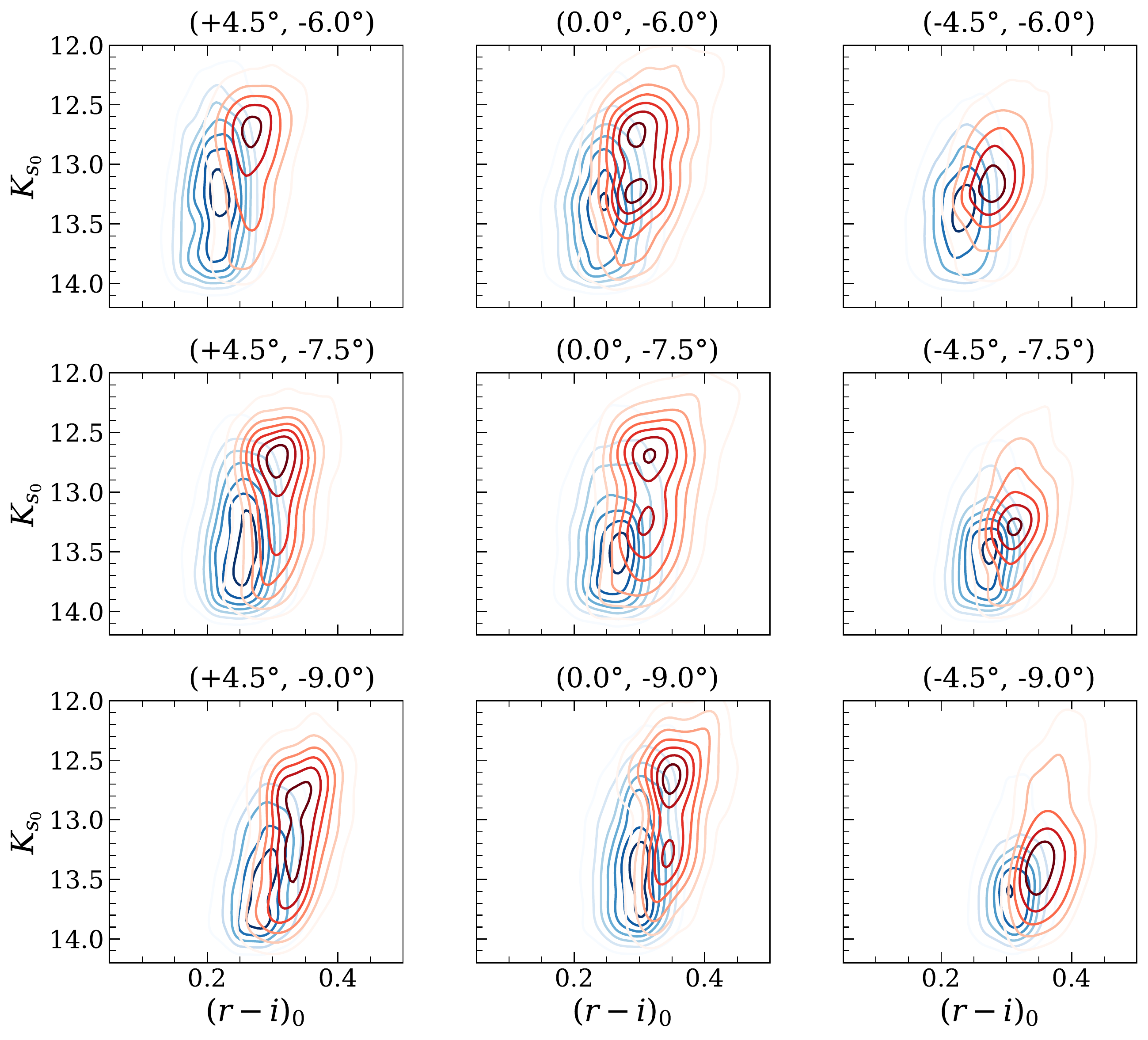}
   \caption{Same as Figure~\ref{app:fig:gi} but for ($r-i$)$_{0}$ color. 
   The bluer and redder stars are overlapped in this color, similar to the case of ($J-K_{s}$)$_{0}$ color (see Figure~\ref{fig:contour_jk}).}
   \label{app:fig:ri}
\end{figure}

\end{appendix}

\end{document}